\documentclass[a4paper,fleqn,usenatbib]{mnras}

\usepackage[T1]{fontenc}
\usepackage{ae,aecompl}
\usepackage{multirow}

\usepackage{graphicx}	
\usepackage{amsmath}	
\usepackage{amssymb}	

\title[Photometric redshifts for radio surveys]{Photometric redshifts for the next generation of deep radio continuum surveys - I: Template fitting}

\author[K. J. Duncan et al.]{Kenneth J Duncan$^{1}$\thanks{E-mail: duncan@strw.leidenuniv.nl},
Michael J. I. Brown$^{2,3}$,
Wendy L. Williams$^{4}$,
Philip N. Best$^{5}$, \newauthor
Veronique Buat$^{6}$,
Denis Burgarella$^{6}$,
Matt J. Jarvis$^{7,8}$,
Katarzyna Ma\l{}ek$^{6,9}$, \newauthor
S. J. Oliver$^{10}$,
Huub J. A. R\"{o}ttgering$^{1}$,
Daniel J. B. Smith$^{4}$
\\
$^{1}$Leiden Observatory, Leiden University, NL-2300 RA Leiden, Netherlands \\
$^{2}$School of Physics, Monash University, Clayton, Victoria 3800, Australia \\
$^{3}$Monash Centre for Astrophysics, Monash University, Clayton, Victoria, 3800, Australia \\
$^{4}$Centre for Astrophysics Research, School of Physics, Astronomy and Mathematics, University of Hertfordshire,\\College Lane, Hatfield AL10 9AB, UK\\
$^{5}$SUPA, Institute for Astronomy, Royal Observatory, Blackford Hill, Edinburgh, EH9 3HJ, UK\\
$^{6}$Aix-Marseille Universit\'{e}, CNRS - LAM (Laboratoire d'Astrophysique de Marseille) UMR 7326, 13388 Marseille, France\\
$^{7}$Astrophysics, University of Oxford, Denys Wilkinson Building, Keble Road, Oxford, OX1 3RH \\
$^{8}$Physics and Astronomy Department, University of the Western Cape, Bellville 7535, South Africa\\
$^{9}$ National Centre for Nuclear Science, ul. Hoza 69, 00-681 Warsaw, Poland\\
$^{10}$Astronomy Centre, Department of Physics and Astronomy, University of Sussex, Falmer, Brighton BN1 9QH, UK\\
}

\pubyear{2016}

\begin{document}
\label{firstpage}
\pagerange{\pageref{firstpage}--\pageref{lastpage}}
\maketitle

\begin{abstract}
We present a study of photometric redshift performance for galaxies and active galactic nuclei detected in deep radio continuum surveys.
Using two multi-wavelength datasets, over the NOAO Deep Wide Field Survey Bo\"{o}tes and COSMOS fields, we assess photometric redshift (photo-z) performance for a sample of $\sim 4,500$ radio continuum sources with spectroscopic redshifts relative to those of $\sim 63,000$ non radio-detected sources in the same fields.
We investigate the performance of three photometric redshift template sets as a function of redshift, radio luminosity and infrared/X-ray properties.
We find that no single template library is able to provide the best performance across all subsets of the radio detected population, with variation in the optimum template set both between subsets and between fields.
Through a hierarchical Bayesian combination of the photo-z estimates from all three template sets, we are able to produce a consensus photo-z estimate which equals or improves upon the performance of any individual template set.
\end{abstract}




\section{Introduction}
Photometric redshifts are a vital tool for estimating the distances to large samples of galaxies observed in extragalactic surveys. At almost all survey scales, from large area surveys such as the Sloan Digital Sky Survey \citep[SDSS;][]{York:2000gn} or the Dark Energy Survey \citep[DES;][]{2005astro.ph.10346T} to deep pencil-beam Hubble Space Telescope (HST) surveys such as CANDELS \citep{2011ApJS..197...35G}, it is impractical to obtain spectroscopic redshifts for more than a small fraction of photometrically detected sources. For the vast majority of sources that are currently detected or will be detected in future photometric surveys, we are therefore reliant on photometric redshift techniques to estimate their distance or extract information about the intrinsic physical properties \citep{2011arXiv1110.3193L}.

While this statement is applicable to photometric surveys across all of the electromagnetic spectrum, the latest generation of deep radio continuum surveys by Square Kilometre Array (SKA) precursors and pathfinders such as the Low Frequency Array \citep[LOFAR;][]{vanHaarlem:2013gi}, the Australian SKA Pathfinder \citep[ASKAP;][]{Johnston:2007ku} and MeerKAT \citep{Booth:2009wx} pose a  new challenge. Probing to unprecedented depths, these surveys will increase the detected population of radio sources by more than an order of magnitude and probe deep into the earliest epochs of galaxy formation and evolution \citep{2010iska.meetE..50R, Jarvis:2012va, Norris:2013jo}.

The population of radio detected sources is itself extremely diverse - with radio emission tracing both black hole accretion in active galactic nuclei (AGN) and star formation activity. With the majority of these sources lacking useful radio morphology information (being unresolved in radio continuum observations), classifying and separating the various sub-populations of radio sources will rely on photometric methods \citep[e.g.][]{Chung:2014it, 2017arXiv170406268C}. Accurate and unbiased photometric redshift estimates for the radio source population will therefore be essential for studying the faint radio population and achieving the scientific goals of these deep radio continuum surveys.

Since the publication of the first widely used photometric redshift (photo-z) estimation tools \citep[e.g.][]{Arnouts:1999gh,Benitez:2000jr,Bolzonella:2000uw}, both the accuracy of photo-z estimates and our understanding of their biases and limitations has significantly improved. The development and testing of photometric redshift techniques has been driven not just by studies of galaxy evolution at high redshifts \citep{Dahlen:2013eu}, but also by the next generation of tomographic weak lensing cosmology surveys \citep{CarrascoKind:2014jg,Sanchez:2014gf}; specifically, the need for computationally fast, accurate and un-biased photometric redshifts for unprecedented samples of galaxies.

Detailed studies have shown that while it is possible to produce accurate photo-zs for X-ray selected AGN \citep{Salvato:2008ef,Salvato:2011dq,Hsu:eu}, care must be taken to correct for the effects of optical variability on photometric data which have been observed over long time periods. Similarly, various studies have been increasingly successful in estimating accurate photometric redshifts for large photometric quasar samples such as the SDSS \citep{York:2000gn}, e.g. \citet{Richards:2001ct}, \citet{Weinstein:2004bh}, \citet{Ball:2008dl}, \citet{Bovy:2012gj}, \citet{Zhang:2013ku} and \citet{Brescia:2013kp}. However, fundamental to all of these efforts is the large representative spectroscopic sample upon which the empirical redshift estimation algorithms are trained.

 Several studies have illustrated that the AGN populations selected at different wavelengths (X-ray, optical, IR, radio) are often distinct, with only some overlap between different selection methods \citep{2009ApJ...696..891H,Kochanek:jy,Chung:2014it}. The optimal photometric redshift techniques and systematics identified for one particular AGN population are therefore not necessarily applicable to an AGN sample selected by other means.

In this paper we aim to quantify some of these systematic effects and find the optimum strategy for estimating accurate photometric redshifts for radio selected populations. 
Specifically, we want to understand how the photometric redshift accuracy of radio sources varies as a function of radio luminosity and redshift. 
Do the current methods and optimization strategies developed for `normal' galaxies or other AGN populations in optical surveys extend to radio selected galaxies? 
Finally, based on the results of these tests, we wish to construct an optimised method which can then be applied successfully to other survey fields in preparation for the next generations of radio continuum surveys \citep[e.g. LOFAR/MIGHTEE:][]{2010iska.meetE..50R, Jarvis:2012va} and the millions of radio sources they will detect \citep{Shimwell:2017ch}.

The paper is structured as follows: Section~\ref{sec:data} outlines the multi-wavelength datasets used in this analysis, including details of optical data used for the photometric redshift estimates and the corresponding radio continuum and spectroscopic redshift datasets.
Section~\ref{sec:photzmethod} then describes how the individual photometric redshift estimates used in this comparison were determined and the choice of software, templates and settings used.
Section~\ref{sec:results} presents the detailed comparison and analysis of these photo-z methods in the context of deep radio continuum surveys.
In Section~\ref{sec:hbmethod} we outline the improved photometric redshift method devised for the LOFAR survey.
Section~\ref{sec:discussion} presents a discussion of the results presented in Section~\ref{sec:results} and their implications for future galaxy evolution and cosmology studies with the forthcoming generation of radio continuum surveys.
Finally, Section~\ref{sec:summary} presents our summary and conclusions. Throughout this paper, all magnitudes are quoted in the AB system \citep{1983ApJ...266..713O} unless otherwise stated. We also assume a $\Lambda$-CDM cosmology with $H_{0} = 70$ kms$^{-1}$Mpc$^{-1}$, $\Omega_{m}=0.3$ and $\Omega_{\Lambda}=0.7$.

\section{Data}\label{sec:data}
To maximise the parameter space explored in this analysis, we make use of two complementary datasets. Firstly, we make use of the extensive multi-wavelength data over the large $\sim9$ deg$^{2}$  NOAO Deep Wide Field Survey in Bo\"{o}tes \citep[NDWFS:][]{Jannuzi:1999wu}. Secondly, we also include data from the COSMOS field which extends to significantly fainter depths across all wavelengths but over a smaller $\sim2$ deg$^{2}$ area.

\subsection{Wide' field - Bo\"{o}tes Field}\label{sec:bootes-photometry}

\subsubsection{Optical photometry}\label{sec:bootes-photometry}
The Bo\"{o}tes photometry used in this study is taken from the PSF matched photometry catalogs of available imaging data in the NDWFS \citep{Brown:2007ca, Brown:2008cu}. The full catalog covers a wide range of wavelengths, spanning from 0.15 to 24~$\mu$m.

The photometry included in the subsequent analysis is based primarily on the deep optical imaging in $B_{W}$, $R$ and $I$-bands from \citet{Jannuzi:1999wu}. At optical wavelengths there is also additional $z$ band coverage from the \emph{zBo\"{o}tes} survey \citep{Cool:2007jg}. Near-infrared observations of the field are provided by NEWFIRM observations at $J$, $H$ and $K_{s}$ \citep{Gonzalez:2010uw}.

Filling in two critical wavelength ranges not previously covered by the existing NOAO Bo\"{o}tes data is additional imaging in the $U_{\text{spec}}$ ($\lambda_{0} = 3590 \AA$) and $y$ ($\lambda_{0} = 9840 \AA$) bands from the Large Binocular Telescope \citep{Bian:2013jk}, covering the full NDWFS observational footprint.
Finally, IRAC observations \citep{Fazio:2004eb} at 3.6, 4.5, 5.8 and 8 $\mu m$ are provided by the \emph{Spitzer} Deep Wide Field Survey (SDWFS, \citealt{Ashby:2009iu}).

Although the available GALEX NUV data cover a significant fraction of the NDWFS field and reach depths comparable to the NOAO $B_{W}$ data, the large point-spread function (PSF) with full-width half maximum (FWHM) equal to $\sim 4.9\arcsec$ could result in significantly increased source confusion relative to the other bands used in the catalog.
As such, the NUV data were not included for the purposes of photometric redshift estimation.\footnote{Initial tests with \textsc{eazy} also found including the NUV data made no appreciable improvement.}

Finally, we also include the \emph{u, g, r, i} and \emph{z} imaging from SDSS \citep{2015ApJS..219...12A}. 
Although the limiting magnitudes reached by the SDSS photometry are not as faint as the NDWFS optical dataset at comparable wavelengths, the different central wavelengths of the SDSS filters provide valuable additional colour information for bright sources and are therefore worth including.

The matched aperture photometry in the catalogs produced by \citeauthor{Brown:2008cu} are based on detections in the NOAO $I$ band image as measured by \textsc{SExtractor} \citep{Bertin:1996hf}.
Forced aperture photometry was then performed on each of the available UV to infrared images for a range of aperture sizes.
The optical/near-infrared images were all first gridded to a common pixel scale and smoothed to a matched PSF.
The common PSF chosen was that of a Moffat profile with $\beta = 2.5$ and a FWHM equal to $1.35\arcsec$ for the $B_{W}$, $R$, $I$, $Y$, $H$ and $Ks$ filters and a larger $1.6\arcsec$ for $u$, $z$ and $J$.

For the matched catalog for photometric redshift estimation, we use fluxes in 3$\arcsec$ apertures for all optical/near-IR bands and 4$\arcsec$ for the IRAC bands.
These aperture fluxes were then corrected to total fluxes using the aperture corrections based on the $1.35$ or $ 1.6 \arcsec$ Moffat profiles or the corresponding IRAC PSF curves of growth.
Tests performed using 2, 3 and 4$\arcsec$ apertures for the optical bands indicate that for Bo\"{o}tes the 3 and 4$\arcsec$ aperture-based photometry perform almost identically for photometric redshift estimation while the 2$\arcsec$-based photometry performed significantly worse.
The choice of 3$\arcsec$ over 4$\arcsec$ apertures is based solely on consistency with the `Deep' data presented in the following sub-section.
 
\subsubsection{Spectroscopic redshifts}\label{sec:bootes-specz}
Spectroscopic redshifts for sources in Bo\"{o}tes are taken from a compilation of observations within the field (Brown, \emph{priv. communication}). 
The majority of redshifts within the sample come from the AGN and Galaxy Evolution Survey \citep[AGES;][]{Kochanek:jy} spectroscopic survey, with additional samples provided by numerous follow-up surveys in the field including \citet{2012ApJ...758L..31L,2013ApJ...771...25L,2014ApJ...796..126L}, \citet{2012ApJ...753..164S}, \citet{2012ApJ...756..115Z,2013ApJ...779..137Z}, \citet{2016ApJ...823...11D} and Hickox, R. C. et al (priv. communication).

The spectroscopic redshift catalog was matched to the combined multi-wavelength catalog based on their quoted physical coordinates in the two respective catalogs and using a maximum separation of 1\arcsec.
In total, the combined sample consists of 22830 redshifts over the range $0 < z < 6.12$, with 88\% of these at $z < 1$.

It is important to raise a caveat to the analysis in the following sections, namely that while the spectroscopic sample used here represents one of best available in the literature and includes a diverse range of galaxy types, it may still not be fully representative of the radio source population.
As with any non spectroscopically complete sample, the subset of sources with available spectroscopic redshifts represents a somewhat biased sample with respect to both the overall photometric sample and the radio selected galaxy population.
In particular, low excitation radio galaxies (LERGS) may be under-represented within the spectroscopic sample due to the lack of strong emission lines available for redshift estimation.

\subsubsection{Radio fluxes}\label{sec:bootes-radio}
Radio observations for the Bo\"{o}tes field are taken from new LOw Frequency ARray \citep[LOFAR;][]{vanHaarlem:2013gi} observations presented in \citet{Williams:2016bf}. Full details of the radio data and  reduction are presented in \citet{Williams:2016bf}, including details of the methods used during calibration and imaging to correct for direction-dependent effects
(DDEs) caused by the ionosphere and the LOFAR phased array beam.

In summary, the observations consist of 8 hr of data taken with the LOFAR High Band Antennae (HBA) and covering the frequency range 130-169 MHz, with a central frequency of $\approx150$ MHz.
The resulting image covers 19 deg$^2$, with a rms noise of $\approx 120 - 150 \mu\textup{Jy~beam}^{-1}$ and resolution of $5.6\times7.4$ arcsec \citep[see][for details on the source extraction and catalog properties]{Williams:2016bf}.

Within the LOFAR field of view, the final source catalog contains a total of 6267 separate $5\sigma$ radio sources. Of these sources, 3902 fall within the boundaries of the $I$-band optical imaging and can therefore be matched to the optical catalogs.
Matches between the LOFAR radio observations and the optical catalog were estimated using a multi-step likelihood ratio technique. Full details of the visual classifications, radio positions and likelihood ratio technique are presented in Williams et al. (in prep). However, the key steps are as follows.
Firstly, radio sources were visually classified into distinct morphological classes.
Next, optical counterparts for each radio source are determined through a likelihood ratio technique based on the positions, positional uncertainty and brightness of the optical and radio sources (with the radio centroid position and uncertainty dependent on the radio morphology classification).
For the small subset of large extended sources where the automated likelihood ratio matching technique cannot be applied, matches were determined individually based on source morphology and visual comparison with the optical imaging.

The cross-matching process yields a total of 2971 matches to sources within the full list of sources within the \citet{Brown:2007ca} optical catalog. However, of these 2971 matches, 578 are matches to optical sources which are flagged as potentially being affected by bright stars/extended sources or are on chip edges.
Of the $\sim 1000$ sources which lie within the $I$-band optical footprint for which no reliable counterpart could be found, a large fraction represent faint un-resolved radio sources for the optical counterpart is too faint.
These sources may be optically faint either due to being at high redshift or as a result of having intrinsically red SEDs.
The Bo\"{o}tes sample used in this analysis is potentially biased against radio sources with optically faint counterparts.
However, thanks to the deep near-IR imaging which forms the basis of the `Deep' COSMOS field we are still able to explore the photo-z properties for these sources.

\subsection{`Deep' field - COSMOS}

\subsubsection{Optical photometry}\label{sec:cosmos-photometry}
The optical/near-IR data used in the COSMOS field are taken from the C OSMOS2015 catalog presented in \citet{Laigle:2016ku}. \citet{Laigle:2016ku} outline fully the details of the optical dataset, including data homogenisation, source detection and extraction. 
We therefore refer the interested reader to said paper for further detail.

For the analysis in this paper, we use the seven optical broad bands ($B$, $V$, $g$, $r$, $i$, $z^{+}$/$z^{++}$), 12 medium bands ($IA427$, $IA464$, $IA484$, $IA505$, $IA527$, $IA574$, $IA624$, $IA679$, $IA709$, $IA738$, $IA767$, and $IA827$) and two narrow bands ($NB711$,$NB816$) taken with the Subaru Suprime-Cam \citep{Capak:2007gb,Taniguchi:2015ey}. 
Also included at optical wavelengths are the $u^{*}$-band data from the Canada-France-Hawaii Telescope (CFHT/MegaCam) as well as Y-band data taken with the Subaru Hyper-Suprime-Cam.
As with the Bo\"{o}tes field, we do not include GALEX UV data in the fitting.
At longer wavelengths we include the UltraVISTA $YJHK_{s}$-band data \citep{McCracken:2012gd} and 3.6, 4.5, 5.8 and 8\micron~ \emph{Spitzer}-IRAC bands.
We make use of the aperture corrected 3\arcsec~ flux estimates for all optical to near-IR bands in combination with the deconfused IRAC photometry as outlined in \citet{Laigle:2016ku}.

\subsubsection{Spectroscopic redshifts}\label{sec:cosmos-specz}
Spectroscopic redshifts for the COSMOS field were taken from the sample of redshifts compiled by and for the Herschel Extragalactic Legacy Project \citep[HELP;][PI: S. Oliver]{2016ASSP...42...71V}.\footnote{The goal of HELP is to produce a comprehensive panchromatic dataset for studying the galaxy population at high redshift - assembling multi-wavelength data and derived galaxy properties over the $\sim1200$~deg$^2$ surveyed by the Herschel Space Observatory.}
The compilation includes the large number of publicly available redshifts in the field \citep[see][and references therein]{Laigle:2016ku} and a small number of currently unpublished samples.
In total, the sample comprises 44,875 sources extending to $z > 6$ and with $\sim12,000$ sources at $z > 1$.

Thanks to the optical depths probed by both the photometric and spectroscopic data available in the COSMOS field, the `Deep' spectroscopic samples a range of galaxy types magnitudes, redshifts which may be missing from the `Wide' sample.
Despite this, the subset of radio detected galaxies with available spectroscopic redshifts may still be biased against towards brighter sources and populations with higher spectroscopic success rates. 

\subsubsection{Radio fluxes}\label{sec:cosmos-radio}
Radio observations for the COSMOS field were taken from the recently released deep VLA observations presented in \citet{2017arXiv170309713S}; the VLA-COSMOS 3 GHz Large Project.
Reaching a median rms $\approx 2.3\mu\textup{Jy/beam}$ over the COSMOS field with at a resolution of $0.75\arcsec$, these observations represent the deepest currently available deep extragalactic radio survey covering a representative volume.
Radio sources from the \citet{2017arXiv170309713S} catalog were matched to their optical counterparts based on the optical matches to \citet{Laigle:2016ku} provided in the companion paper \citet{2017arXiv170309719S}.
Within the spectroscopic redshift subsample there are a total of 3400 radio detected sources.
While a comparison of the difference in the redshift distribution and source types between a 150 MHz and 3GHz selected survey may be of scientific interest, it is a topic which we do not intend to address here.
To facilitate direct comparison with the LOFAR 150MHz fluxes, we convert the observed 3 GHz fluxes to estimated 150MHz fluxes assuming a median 3000 to 150MHz spectral slope of $\alpha = -0.7$ \citep{2017arXiv170309713S,2017arXiv170406268C}.

\subsection{Flagging of known X-ray sources and known IR/Optical AGN}\label{sec:mw_agn}
In deep radio continuum surveys, the radio detected population includes a very diverse range of sources, ranging from rapidly star-forming galaxies to radio quiet quasars and massive elliptical galaxies hosting luminous radio AGN.
To fully characterise the diverse radio population and to facilitate comparison between the radio population and other AGN selection methods, we classify all sources in the spectroscopic comparison samples using the following additional criteria:
\begin{itemize}
\item \emph{Infrared AGN} are identified using the updated IR colour criteria presented in \citet{Donley:2012ji}.
In addition to the colour criteria outlined by \citet{Donley:2012ji}, we split the IR AGN sample into two subsets based on their signal to noise in the IRAC 5.6 and 8$\mu$m bands.
To be selected as a candidate IR AGN, we require that all sources have $S/N > 5$ at 3.5 and 4.6$\mu$m and $S/N > 2$ at 5.6 and 8$\mu$m.
The subset of robust AGN sources is then based on a stricter criteria of $S/N > 5$ at 5.6 and 8$\mu$m.

\item \emph{X-ray} selected sources in the Bo\"{o}tes field were identified by cross-matching the positions of sources in our catalog with the X-B\"{o}otes \emph{Chandra} survey of NDWFS \citep{Kenter:2005gj}.
We matched the x-ray sources through a simple nearest-neighbour match between the optical photometry catalog used in this analysis and the position of most-likely \textit{optical} counterpart for each x-ray source presented in \citep{Brand:2006iv}.
In COSMOS, we make use of the compilation of matched x-ray data presented in \citet{Laigle:2016ku} and the corresponding papers detailing the X-ray sources and optical cross-matching \citep{2016ApJ...819...62C,2016ApJ...817...34M}.

For both fields we calculate the x-ray-to-optical flux ratio, $X/O = \log_{10}(f_{X}/f_{\textup{opt}})$, based on the $i^{+}$ or $I$ band magnitude following \citet{Salvato:2011dq} and \citet{Brand:2006iv} respectively. 
To be selected as an X-ray AGN, we require that an x-ray source have $X/O > -1$ or an x-ray hardness ratio $> 0.8$ \citep{2004AJ....128.2048B}.

\item Bright, known \emph{Optical AGN} were also identified through two additional selection criteria. Where available, any sources which have been spectroscopically classified as AGN are flagged. Secondly, we also cross-match the optical catalogs with the Million Quasar Catalog compilation of optical AGN, primarily based on SDSS  \citep{2015ApJS..219...12A} and other literature catalogs \citep{2015PASA...32...10F}. 
Objects in the million quasar catalog were cross-matched to the photometric catalogs using a simple nearest neighbour match in RA and declination and allowing a maximum separation of 1\arcsec.
Simulations using randomised positions indicate that at $1\arcsec$ separation, the chance of a spurious match with an object in the optical catalog is less than 5\%.
While this value is relatively high due to the depth of the optical catalogs, the actual median separation between matches is $\approx 0.2\arcsec$ and are highly unlikely to be spurious.
Visual inspection of the quasar catalog sources with no optical counterpart in our catalog indicates that the majority fall within masked regions of the optical catalog (e.g. around bright stars and artefacts) and thus are not expected to have a match.
\end{itemize}

\begin{figure}
\centering
	\includegraphics[width=0.5\columnwidth]{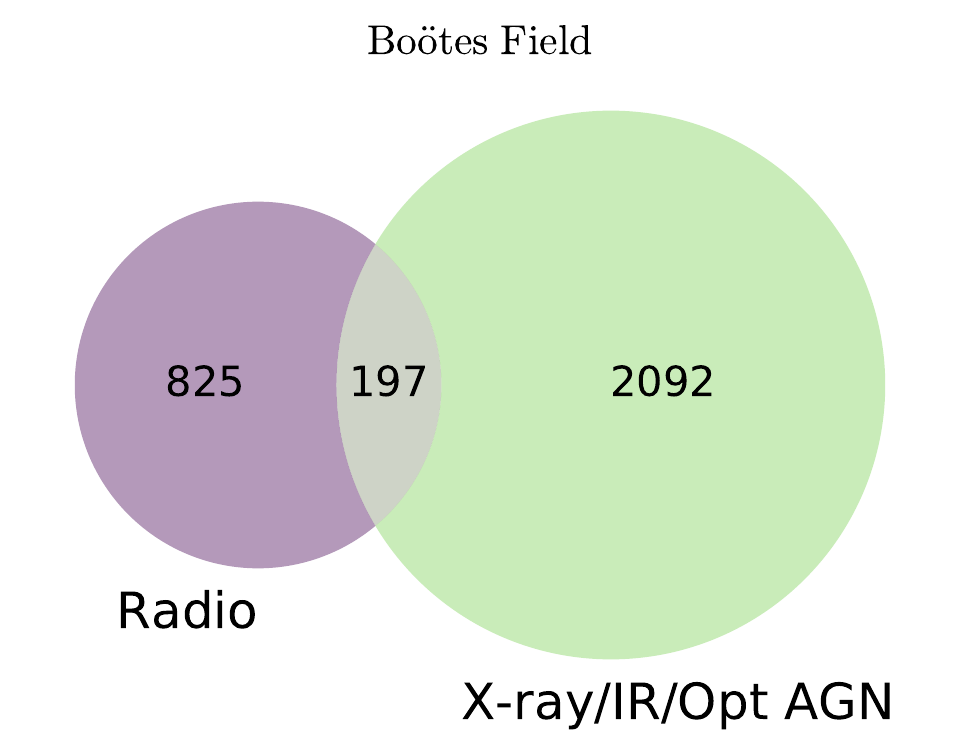}\includegraphics[width=0.5\columnwidth]{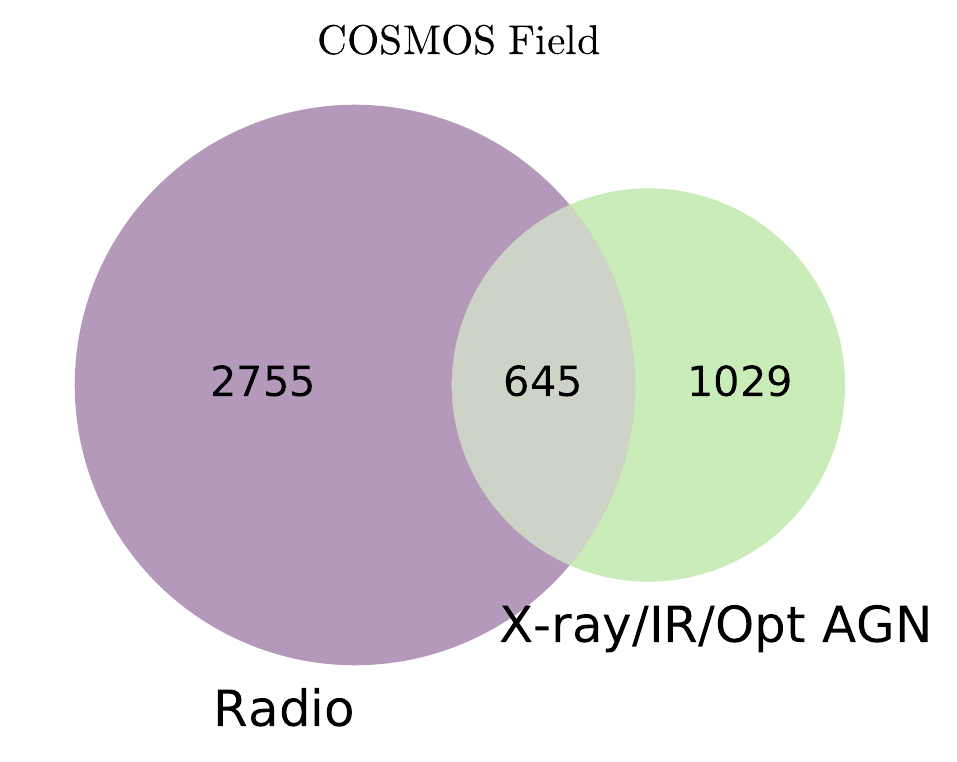}
  \caption{Multi-wavelength classifications of the sources in the full spectroscopic redshift samples. The `Radio' and  `X-ray/IR/Opt AGN' subsets correspond respectively to radio detected sources and identified X-ray sources and optical/spectroscopic/infra-red selected AGN (see Section~\ref{sec:mw_agn}). As illustrated in previous studies, the X-ray, IR AGN and radio source population are largely distinct populations with only partial overlap.}
  \label{fig:venn_all}
\end{figure}

\begin{figure}
\centering
	\includegraphics[width=0.5\columnwidth]{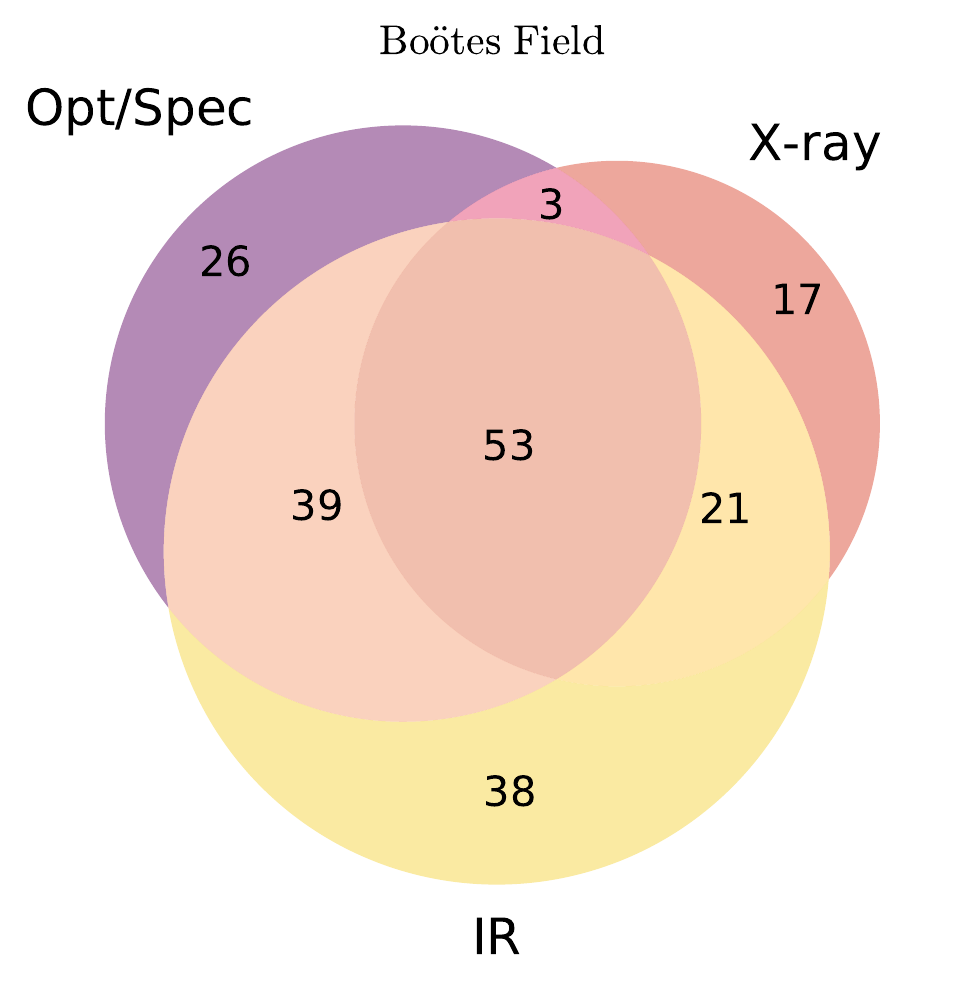}\includegraphics[width=0.5\columnwidth]{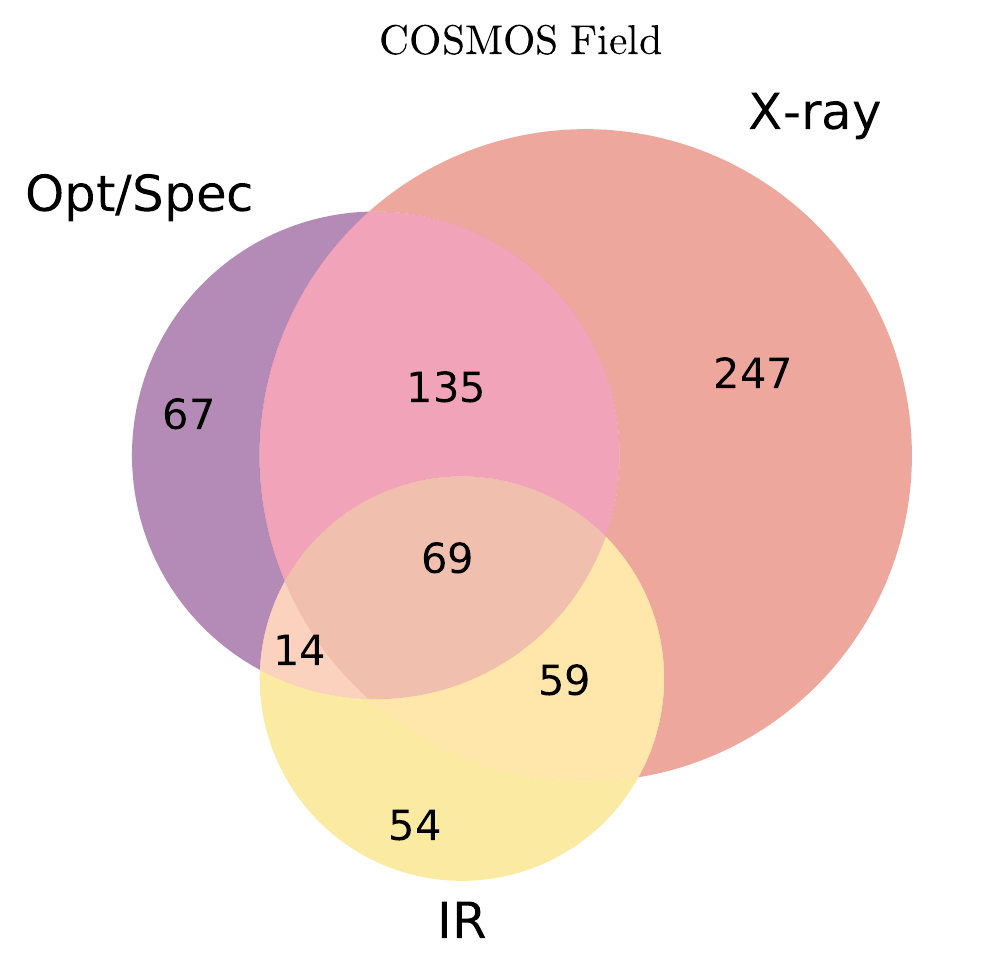}
  \caption{Multi-wavelength classifications of the radio detected sources within the spectroscopic redshift sample. For the 214 and 711 radio detected sources in the Bo\"{o}tes and COSMOS fields respectively, subsets which satisfy the X-ray, Optical (`Opt/Spec') and IR AGN criteria  as defined in Section~\ref{sec:mw_agn}.}
  \label{fig:venn_radio}
\end{figure}

In Fig.~\ref{fig:venn_all} we show the relative numbers of radio detected sources and sources which satisfy any of the X-ray/Optical IR AGN critera within the full spectroscopic subsets. In both the Bo\"{o}tes and COSMOS spectroscopic samples, there are large numbers of radio or X-ray detected sources, as well as large numbers of sources classified as IR AGN. For the Bo\"{o}tes field, the large number of IR AGN is due to the specific selection criteria targeting these sources within the AGES spectroscopic survey \citep{Kochanek:jy}.

\begin{figure}
\centering
	\includegraphics[width=0.5\columnwidth]{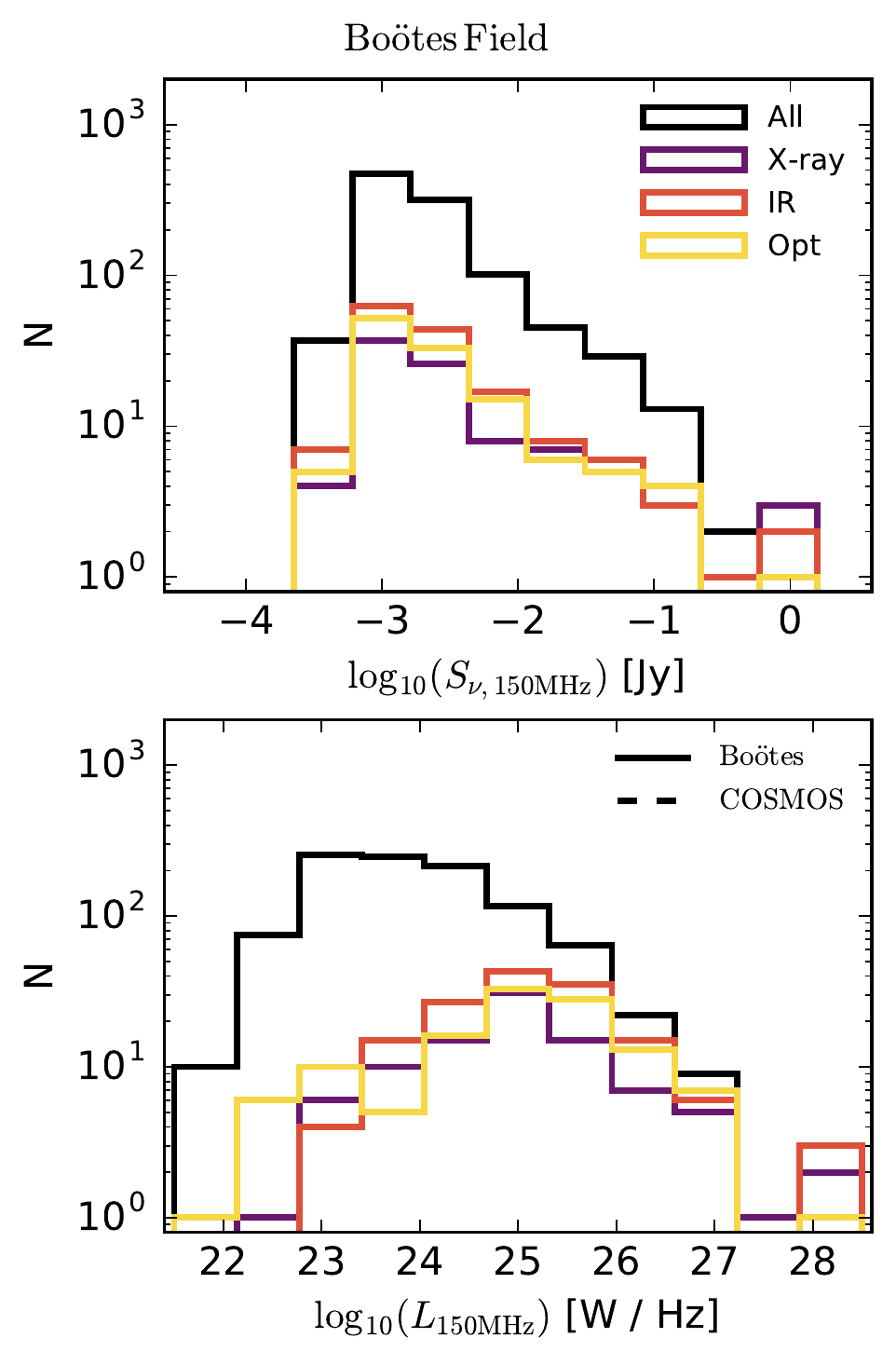}\includegraphics[width=0.5\columnwidth]{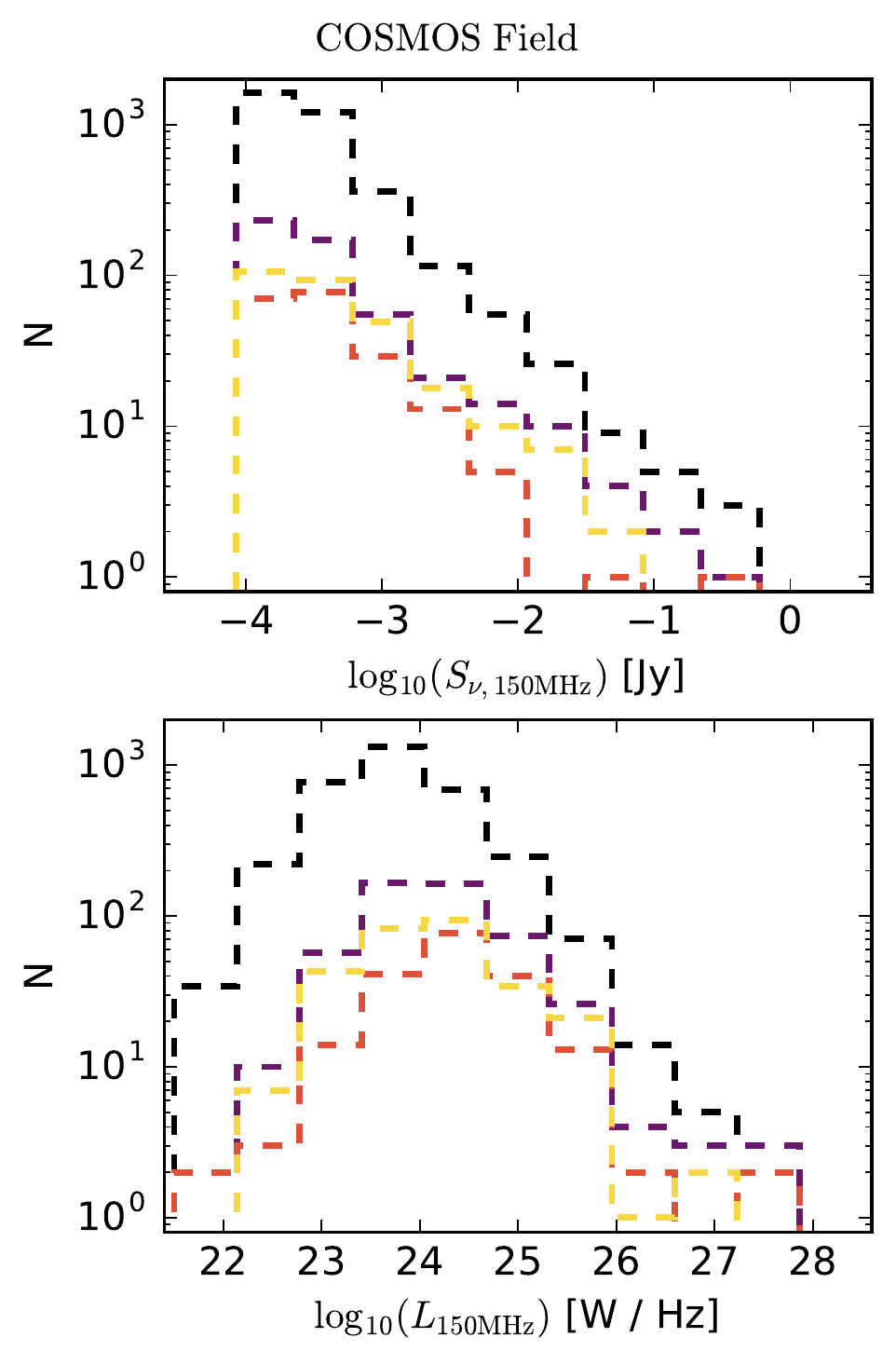}

  \caption{Flux and luminosity distributions (top and bottom rows respectively) of the radio detected sources within the Bo\"{o}tes and COSMOS (left and right columns)  spectroscopic sample used in this analysis. Plotted in both pairs of histograms are the fluxes (or luminosities) of all radio detected sources (black lines) as well as those which are also X-ray sources \citep[][purple lines]{Brand:2006iv}, infrared AGN following the criteria of \citet[][orange lines]{Donley:2012ji}, or optical/spectroscopically identified AGN (yellow lines).}
  \label{fig:fluxlum_hists}
\end{figure}

Within the subset of radio detected sources itself there is a clear diversity in the nature of sources. In Fig.~\ref{fig:venn_radio} we show the multi-wavelength classifications of the respective radio samples.

\begin{figure}
\centering
	\includegraphics[width=0.98\columnwidth]{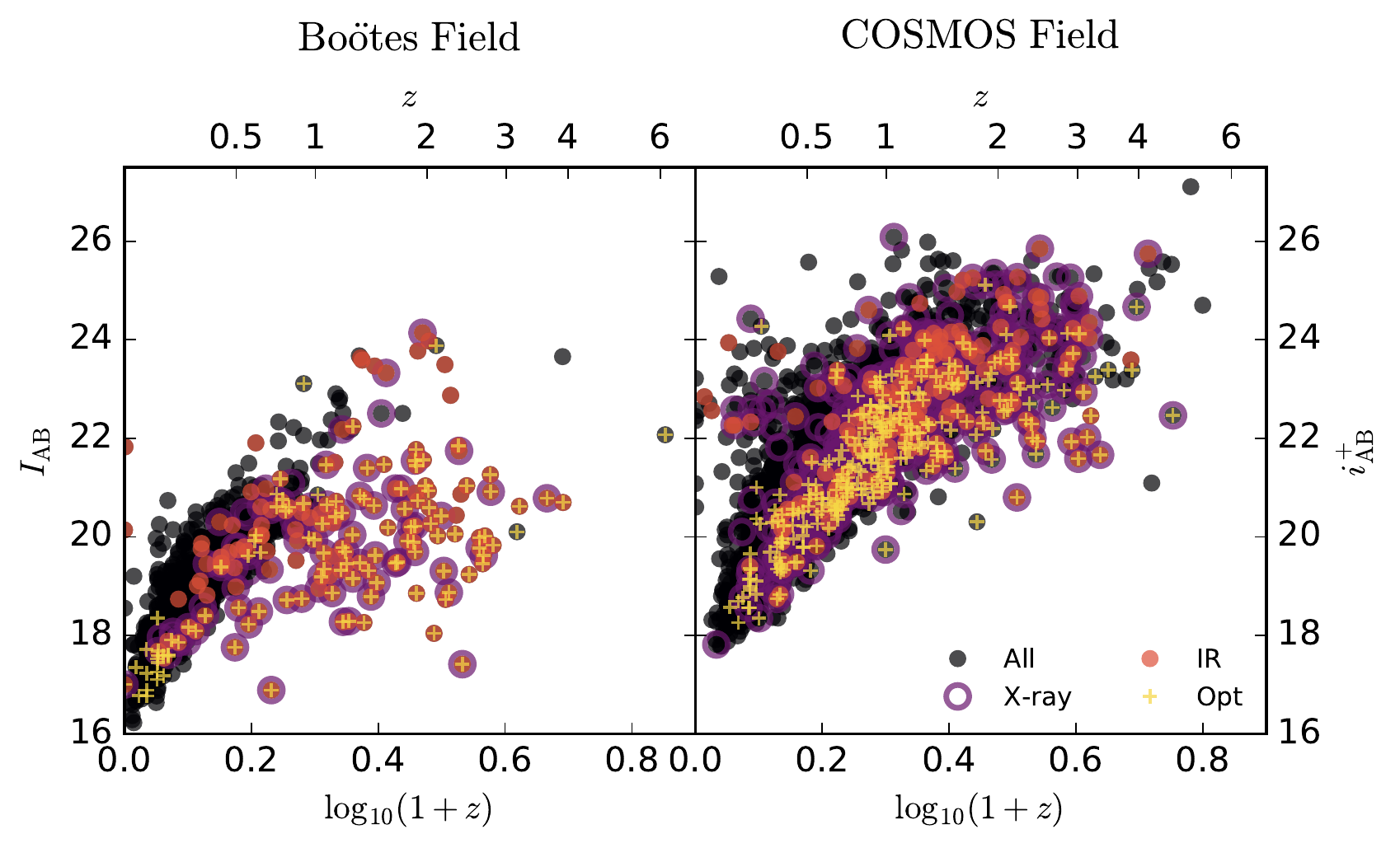}
  \caption{Optical magnitude vs redshift distributions of the radio detected sources within the Bo\"{o}tes and COSMOS (left and right panels respectively) spectroscopic sample used in this analysis. Plotted in both plots are the $I$ ($i^{+}$) magnitude and spectroscopic redshift of all radio detected sources (black circles). For each source, additional markers are added to illustrate whether it is X-ray detected (purple ring), an infrared AGN following the criteria of \citet[][orange circle]{Donley:2012ji}, or an optical/spectroscopically identified AGN (light yellow crosses).}
  \label{fig:mag_vs_z}
\end{figure}

Inspecting the radio flux and luminosity distributions of the two samples (Fig.~\ref{fig:fluxlum_hists}) reveals that the X-ray detected sources and IR AGN typically have a higher radio luminosity than the sample median - in line with the expected dominance of AGN at $L_{150\textup{MHz}} \gtrsim10^{24} ~ \textup{W/Hz}$ \citep{Jarvis:2004hd,Padovani:2016fo}.
However as seen in Fig.~\ref{fig:venn_radio}, of the most radio luminous sources, e.g. $L_{150\textup{MHz}} > 10^{25} ~ \textup{W/Hz}$, only $\sim 40 - 50\%$ also satisfy another AGN selection criterion.
Of all X-ray and IR AGN sources in our samples, we note that $\approx 10-20\%$ are radio detected, broadly consistent with the measured radio-loud fraction of optical quasars \citep{Jiang:2013vf}.

Finally, to illustrate the magnitude and redshift parameter space probed by our spectroscopic redshift comparison samples, in Fig.~\ref{fig:mag_vs_z}, we plot the apparent $I$($i^{+}$) band magnitudes and redshifts of the radio detected populations.
By construction, the `Deep' COSMOS sample probes to significantly fainter magnitudes than the wide area Bo\"{o}tes sample.
Between the two samples we are able to sample a wide range of magnitudes in the redshift range $0 < z < 3$ ($0 < \log_{10}(1+z) \lesssim 0.6$).
We also caution that due to the nature of the AGES spectroscopic survey selection criteria \citep{Kochanek:jy}, the majority of spectroscopic redshifts at $z > 1$ in the Bo\"{o}tes field are known AGN.
Conclusions on the photo-zs for sources at $z > 1$ will therefore largely be driven by the less biased COSMOS sample.

\section{Photometric redshift methodology}\label{sec:photzmethod}
Photometric redshift estimation techniques fall broadly into two distinct categories. Firstly, one can use redshifted empirical or model template sets fitted to the model photometry through $\chi^{2}$-minimisation or maximum likelihood techniques \citep[e.g.][]{Arnouts:1999gh,Bolzonella:2000uw,Benitez:2000jr,Brammer:2008gn}.
Alternatively, one can take a representative training set of objects that has known spectroscopic redshifts and use any of a wide variety of supervised or un-supervised machine learning algorithms to estimate the redshifts for the sample of galaxies for which the redshift is unknown \citep[e.g.][]{Collister:2004fx,Brodwin:2006dp,CarrascoKind:2013kd,CarrascoKind:2014gb, 2016MNRAS.455.2387A,2016MNRAS.462..726A}.

In recent years, empirical methods based on training sets have been shown to produce redshift estimates that can have lower scatter and outlier fractions than template-based methods \citep{Sanchez:2014gf,CarrascoKind:2014jg}. Furthermore, because the computationally expensive training step only occurs once these methods can also be significantly faster than template fitting when applied to very large datasets.

However, the drawback of training sample methods is that they are very dependent on the parameter space covered by the training sample and its overall representativeness of the sample being fitted \citep{Beck:2017wq}. While template-fitting methods do benefit from additional optimisation through spectroscopic training samples \citep[Section~\ref{sec:zeropoints}, see also][]{Hildebrandt:2010de,Dahlen:2013eu}, they can be applied effectively with no prior redshift knowledge and tested without spectroscopic samples for comparison \citep{Quadri:2010bv}.

Fully representative training samples for the rare sources of interest are not yet readily available for many different fields. 
Contributing to this problem is the inhomogeneous nature of the photometric data both within and across the various deep survey fields. 
While deep spectroscopic samples are available for fields such as COSMOS, the variation in filter coverage between survey fields makes it impractical to fully apply this training sample to other fields.
Given these constraints, we believe that template based photometric redshifts still represent the best starting point when estimating photo-zs for the datasets and science goals of interest.
Future work will explore the application of such empirical photo-z estimates to the widest tiers of the LOFAR survey.
For this study we base the photometric redshift estimates on the \textsc{eazy} photometric redshift software presented in \citet{Brammer:2008gn}. 

As mentioned above, several different template fitting photometric redshift codes have been published and have been widely used in the literature, e.g. \textsc{BPZ} \citep{Benitez:2000jr}, \textsc{LePhare} \citep{Arnouts:1999gh,2005A&A...439..863I} or \textsc{HyperZ} \citep{Bolzonella:2000uw}. 
The key differences in approach (and potential outcomes) between these codes are primarily the choice in default template sets as well as their treatment of redshift priors based on magnitude or spectral type. 
Both of these assumptions can be changed either within \textsc{eazy} itself or in subsequent analysis of its outputs. 
We are therefore confident that our choice of specific photometric redshift code does not strongly bias the results of our analysis and we note that alternative template fitting codes could be used without systematically affecting the results.

\subsection{Template sets}

The three template sets used in this analysis are as follows:
\begin{enumerate}
	\item Default \textsc{eazy} reduced galaxy set (`EAZY'):

\indent The first set used are the updated optimised galaxy template set provided with  \textsc{eazy} and we refer the reader to \citet{Brammer:2008gn} for full details of how these templates were generated.
 In the latest version of the software, this template set has been updated to incorporate nebular emission lines and includes both an additional dusty galaxy template and an extremely blue SED with strong line emission.

Because the \textsc{eazy} template set includes only stellar emission it gives poor fits at wavelengths where the overall emission is typically dominated by non-stellar radiation (e.g. rest-frame mid-infrared; dust emission/PAH features).
To minimise the effect of this potential bias, observed filters with wavelengths greater than that of IRAC channel 2 (4.5\micron) are not included when fitting.

	\item \citet{Salvato:2008ef} `XMM-COSMOS' templates:

\indent Our second set of templates is that presented by \citet{Salvato:2008ef,Salvato:2011dq} in their analysis of photometric redshifts for X-ray AGN.
Based on the templates presented in \citet[][see also references within]{Polletta:2007ha}, this template set includes 30 SEDs and covers a wide range of galaxy spectral types in addition to both AGN and QSO templates.
In contrast to the \textsc{eazy} templates, the XMM-COSMOS templates include both dust continuum and PAH features as well as power-law continuum emission for the appropriate AGN templates.
We therefore do not exclude the IRAC 5.8 and 8.0 $\mu$m photometry when fitting with these templates.

	\item \citet{Brown:2014jd} Atlas of Galaxy SEDs (`Atlas'):

\indent	Finally, we make use of the large atlas of 129 galaxy SED templates presented in \citet[][referred to as `Atlas' hereafter]{Brown:2014jd}.
These templates are based on nearby galaxies and cover a broad range of galaxy spectral types including ellipticals, spirals and luminous infrared galaxies (both starburst and AGN).
Constructed from panchromatic synthetic SED models \citep{daCunha:2008cy} and optical to infrared photometry and spectroscopy, the library has been constructed to minimise systematic errors and span the full gamut of nearby galaxy colours.
As with the template set 2, because the templates include rest-frame mid-infrared spectral and continuum features, IRAC 5.8 and 8.0 $\mu$m photometry were also used when fitting with this library.
\end{enumerate}

\begin{table*}
  \caption{Summary of the three template sets used in the photometric redshift analysis, including details of how the templates were fitted, whether any dust attenuation was applied to the original template sets and whether the template set includes contributions from AGN emission.}
  \label{tab:templates}
  \begin{tabular}{ccccc}
  \hline
  Template Set  & N Templates & Fitting Mode & Dust Attenuation Applied & AGN Included\\
    \hline
    EAZY & \multirow{2}{*}{9} &  \multirow{2}{*}{$N$-linear combinations} &  \multirow{2}{*}{N/A} & \multirow{2}{*}{No} \\
    \citet{Brammer:2008gn} & & & &\\
     & & & &\\
    XMM-COSMOS & \multirow{2}{*}{32} & \multirow{2}{*}{Single Template} & $0 \leq A_{V} \leq 2$, $\delta A_{V} = 0.2$ & \multirow{2}{*}{Yes} \\
    \citet{Salvato:2008ef,Salvato:2011dq}  & & & \citet{2000ApJ...533..682C}, \citet[][SMC]{Pei:1992ey} & \\
     Atlas of SEDs &  \multirow{2}{*}{129} & \multirow{2}{*}{Single Template} & $0 \leq A_{V} \leq 1$, $\delta A_{V} = 0.2$  & \multirow{2}{*}{Yes} \\
    \citet{Brown:2014jd}  & & & \citet{2000ApJ...533..682C} &\\
    & & & &\\
    \hline
     \end{tabular}
\end{table*}

These three template libraries were selected either because of their common use within the literature (EAZY/XMM-COSMOS) or because of their explicit intention to fully represent the range of colours observed in local galaxies (Atlas).
They are however not directly comparable in the intrinsic galaxy types they include and there are some key differences which could affect their potential performance for the radio galaxy population.
As mentioned above, the EAZY template set models only stellar emission and does not include any templates with contributions from AGN.
We may therefore expect the EAZY template set to perform very poorly for galaxies with SEDs which are dominated by AGN components.

In contrast, while the Atlas library does include templates with significant AGN contributions (primarily at longer wavelengths), it does not include any bright optical quasars due to its local galaxy selection.
The XMM-COSMOS library is therefore the only set included in this analysis which includes the full range of optical AGN classes.

\subsection{Photometric zeropoint offsets}\label{sec:zeropoints}
The addition of small magnitude offsets to the observed photometry of some datasets has been shown to improve photometric redshift estimates \citep{Dahlen:2013eu}. While typically small ($\lesssim 10\%$), these additional offsets can often substantially reduce the overall scatter or outlier fractions for photo-z estimates.

To calculate the appropriate photometric offsets we use the commonly followed strategy of fitting the observed SEDs of a subset of galaxies while fixing their redshift to the known spectroscopic redshift.
For a training sample of 80\% of the available spectroscopic sample, the offset for each band is then calculated from the median offset between the observed and fitted flux values for sources with $S/N > 3$ in that band.

To ensure that spurious offsets are not being applied based on a small number of catastrophic failures in the photometry we perform a bootstrap analysis to calculate the scatter in estimated zeropoint offsets.
The zeropoint offset is calculated for 100 iterations of a random subset of $10\%$ of the spectroscopic training sample, with the standard deviation of this distribution then taken as the uncertainty in the zeropoint offset. 
An offset is then only applied to a given band if the offset is significant at the 2$\sigma$ level.

We apply this procedure to each template set individually, with the zeropoint offsets applied in all subsequent analysis steps. Using the remaining 20\% of spectroscopic redshifts as a test sample we are able to verify that for each template set the inclusion of the zeropoint offsets in the fitting produces an overall improvement in the various photometric redshift quality metrics.

Finally, before including the estimated photometric offsets in the fitting process for the full photometric samples we assess any potential adverse effects they could have.
For the two example fields used in this study we find that there is no strong bias in the photometric offsets introduced by the redshift distribution of the spectroscopic sample.
That is to say, applying photometric offsets based on a spectroscopic sample with $\overline{z_{s}} \approx 0.3$ to a sample of photometric galaxies at higher redshift will not strongly bias the resulting redshifts.
Such biases could arise either from aperture effects (due to the larger angular size of nearby galaxies) or from differences in the age-dependent features (e.g. 4000\AA break) in the SEDs; a problem which which may be most acute for the local galaxy based Atlas template library.
However, we find that for the extreme example of applying photometric offsets calculated for a spectroscopic sample at $z\sim0.2$ to a test sample at $z > 1$, the photometric redshift quality of the test sample with the `biased' offsets applied is not significantly worse than when no offsets were applied.

\subsection{Fitting methods}

The EAZY template set is fitted following their intended use, using fits of N-linear combinations of templates and allowing all templates to be included in the fit.

In contrast, the XMM-COSMOS templates are used in a way which best matches their implementation in \textsc{LePhare} \citep{Arnouts:1999gh} and their intended use  \citep{Salvato:2011dq}.
A range of dust attenuation levels ($0 \leq A_{V} \leq 2$) is applied to each of the 32 unique templates, using both the \citet{2000ApJ...533..682C} starburst attenuation law and the \citet{Pei:1992ey} Small Magellanic Cloud (SMC) extinction curve.
The extended set of dust attenuated templates are then fitted using single template mode in \textsc{eazy}.

Due to the large number of unique templates already included (making fits of N-linear combinations impractical), the Atlas template set is fitted in a similar manner to the XMM-COSMOS set.
To allow for finer sampling of the rest-frame UV/optical emission in the empirical Atlas SEDs, we also apply additional dust attenuation to the empirical templates as was done for the XMM-COSMOS set.
Due to the wider range of dust extinction already intrinsic to the empirical templates, we apply a smaller range of additional dust attenuation ($0 \leq A_{V} \leq 1$) and assume only the \citet{2000ApJ...533..682C} starburst attenuation law.
We note that the maximum dust extinction of $A_{V} = 1$ may be unrealistic for some of the galaxy archetypes included in the Atlas library (e.g. blue compact dwarfs), but dust ranges tuned to individual template type is beyond the scope of this work.
As for the XMM-COSMOS template set, the extended Atlas of Galaxy SEDs template set is then fit in single template mode.

For all three template sets, additional rest-frame wavelength dependent flux errors are also included through the \textsc{eazy} template error function \citep[see][]{Brammer:2008gn}.
These errors are added in quadrature to the input photometric errors and vary from $<5\%$ at rest-frame optical wavelengths to $>15\%$ at rest-frame UV and near-IR where template libraries are more poorly constrained.

Finally, although \textsc{eazy} allows for the inclusion of a magnitude dependent prior in the redshift estimation, we choose not to include it at this stage.
A summary of these three different photo-z fitting estimates is presented in Table~\ref{tab:templates} for reference.

\section{Results}\label{sec:results}
To explore the performance of the three template sets on the two spectroscopic samples, we first look at the statistics of the best-fit photometric redshift estimates relative to the measured spectroscopic redshifts.
Within the literature there are a wide range of statistical metrics used to quantify the quality of photometric redshifts \citep[see][]{Dahlen:2013eu,Sanchez:2014gf,CarrascoKind:2014jg}.
In this analysis we choose to adopt a subset of the metrics outlined in \citet{Dahlen:2013eu}, including three measures of the redshift scatter, one measure of the bias and one of the outlier fraction (see Table~\ref{tab:definitions} for details of these definitions and their notation).

We also introduce an additional metric, the continuous ranked probability score (CRPS) and the corresponding mean values for a given sample \citep{Brown:1974td,Matheson:1976he}.
Widely used in meteorology, the CRPS is designed for evaluating probabilistic forecasts. 
We refer to \citet[][see also: \citet{2016arXiv160808016P}]{2001math......1268R} for full details of the metric and its behaviour, but its definition is presented in Table~\ref{tab:definitions} and represents the integral of the absolute difference between the cumulative redshift distributions of the predicted value  ($\textup{CDF}(z)$)  and true values  ($\textup{CDF}_{z_{s}}(z)$: i.e. a Heaviside step function at $z_{s}$).
A key advantage over the more widely used metrics is that CRPS takes into account the full PDF rather than just a simple point value when evaluating a model prediction (i.e. the photometric redshift).

\subsection{Overall photometric redshift accuracy}\label{sec:overall}
\begin{figure}
\centering
\includegraphics[width=0.9\columnwidth]{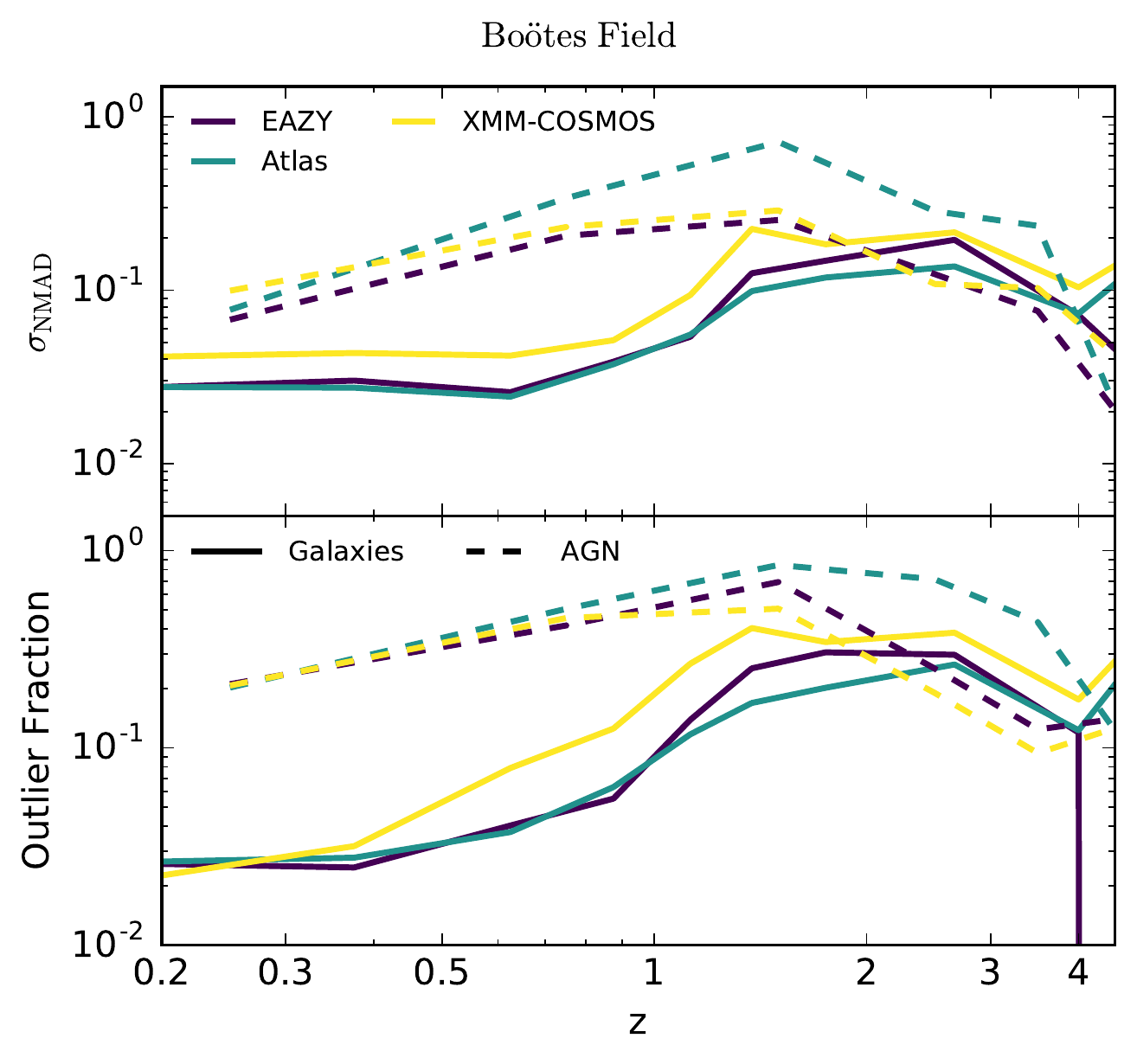}
\includegraphics[width=0.9\columnwidth]{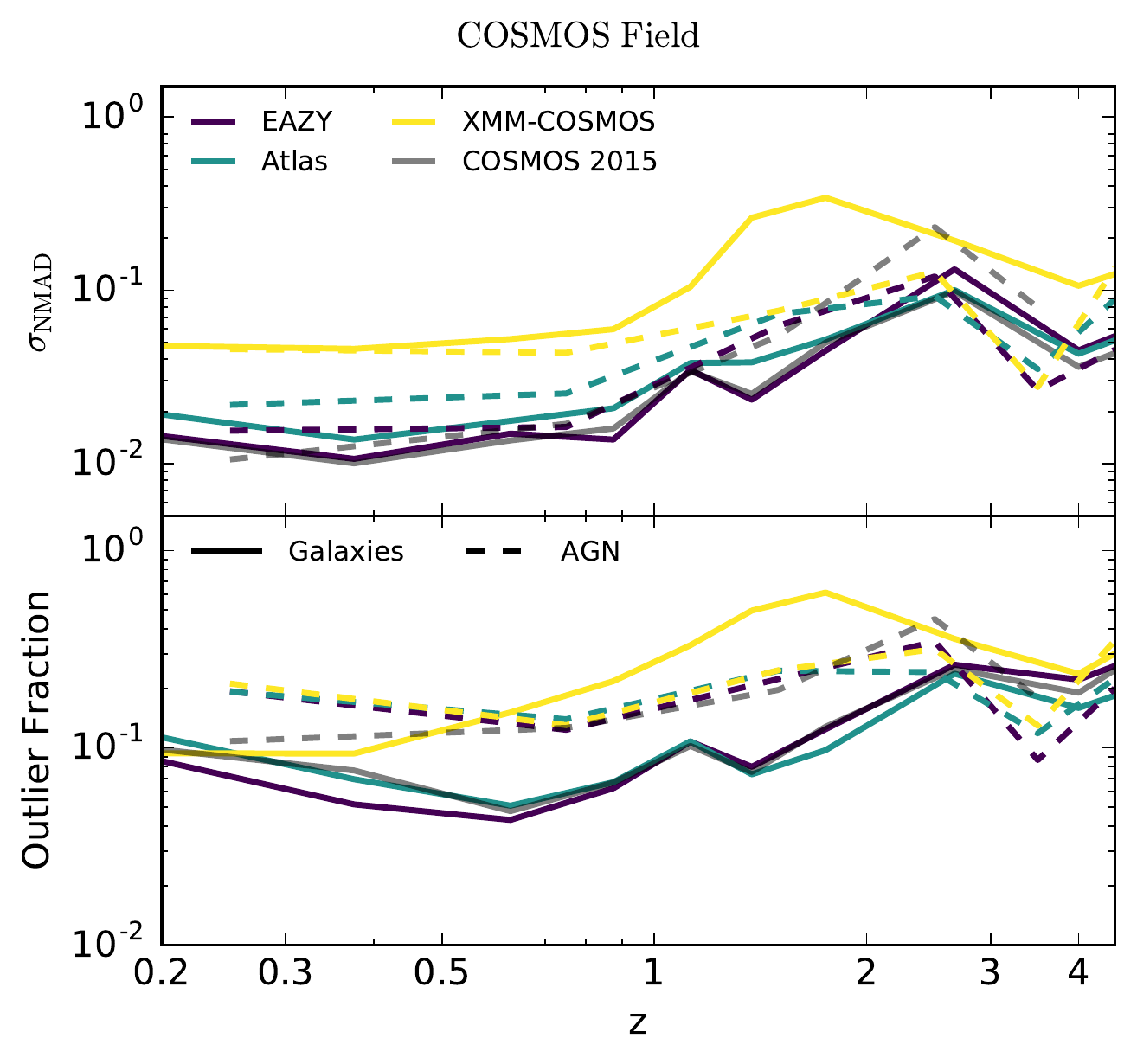}
  \caption{Photometric redshift scatter ($\sigma_{\text{NMAD}}$) and outlier fraction as a function of spectroscopic redshift for the Bo\"{o}tes field (top) and COSMOS fields (bottom) respectively. In both plots, dashed lines show the results for sources which pass any of the X-ray/Optical/IR AGN criteria outlined in Section~\ref{sec:mw_agn} and solid lines show the results for sources which do not satisfy any of these criteria. The `COSMOS 2015' line corresponds to the combined literature COSMOS photometric redshift values from \citet{Laigle:2016ku} and \citet{Marchesi:2016eo}.}
  \label{fig:sigma_vs_z}
\end{figure}

\begin{figure}
\centering
\includegraphics[width=0.9\columnwidth]{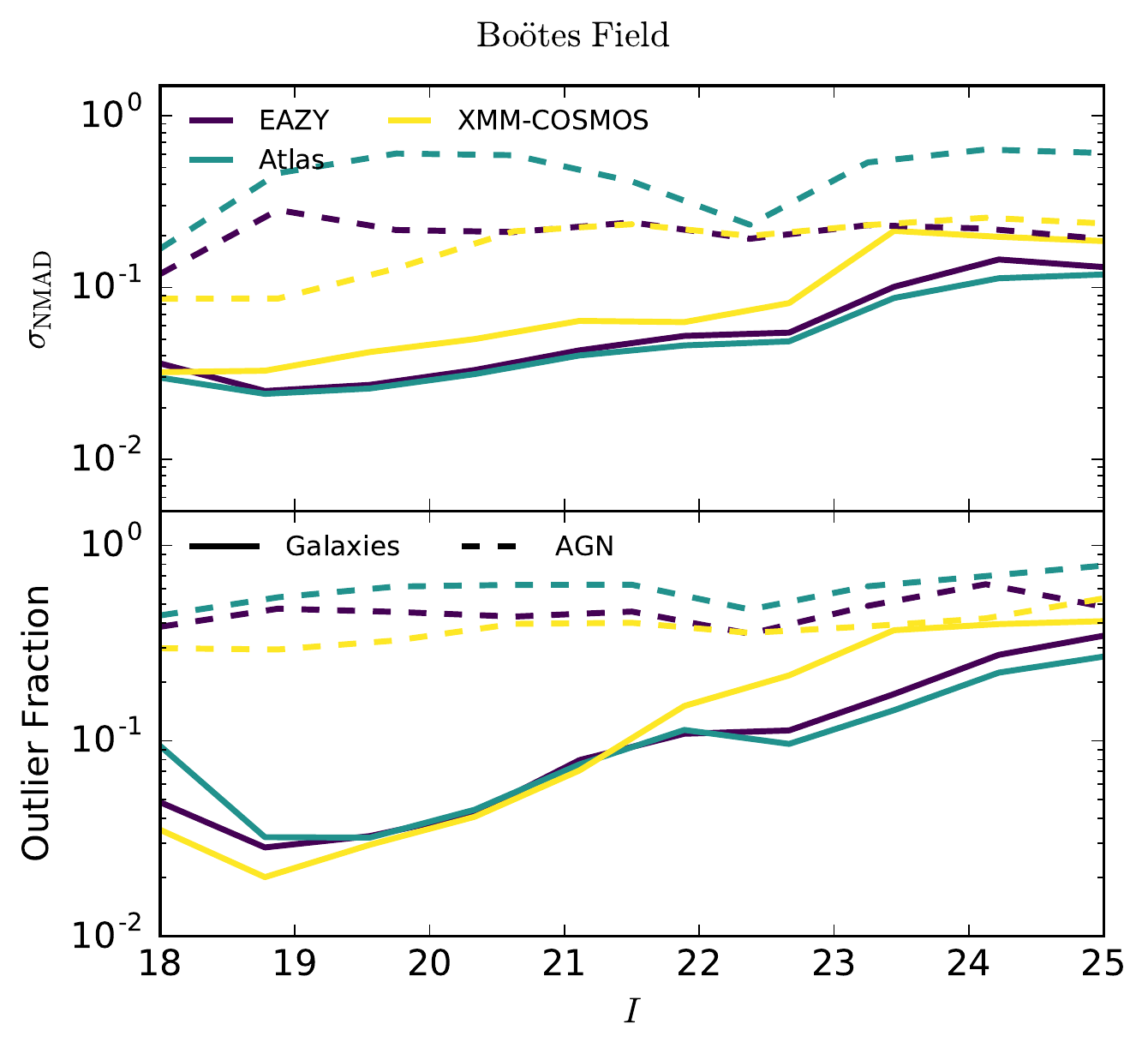}
\includegraphics[width=0.9\columnwidth]{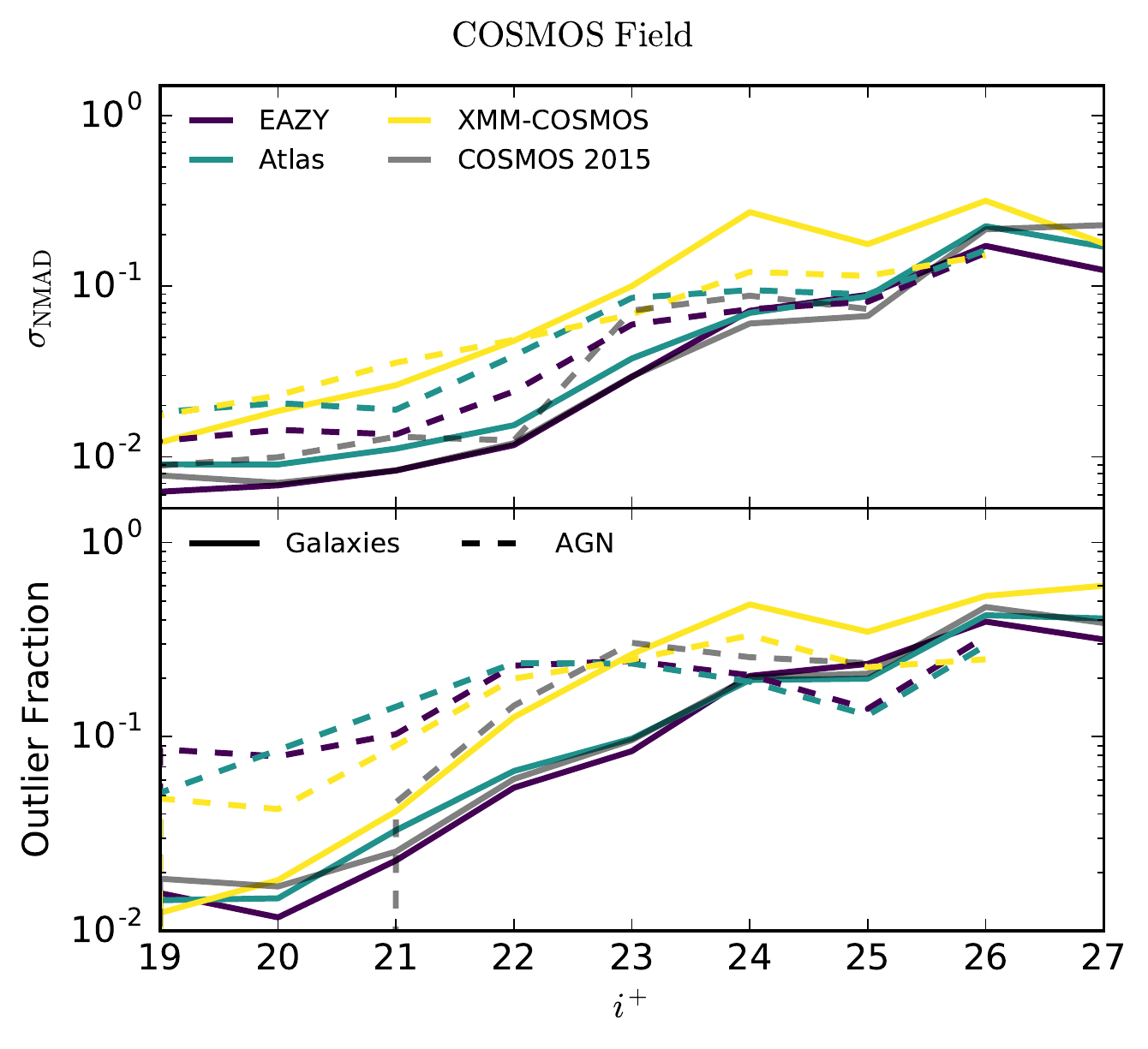}
  \caption{Photometric redshift scatter ($\sigma_{\text{NMAD}}$) and outlier fraction as a function of $I$, or $i^{+}$, magnitude for the Bo\"{o}tes field (top) and COSMOS fields (bottom) respectively. In both plots, dashed lines show the results for sources which pass any of the X-ray/Optical/IR AGN criteria outlined in Section~\ref{sec:mw_agn} and solid lines show the results for sources which do not satisfy any of these criteria. The `COSMOS 2015' line corresponds to the combined literature COSMOS photometric redshift values from \citet{Laigle:2016ku} and \citet{Marchesi:2016eo}.}
  \label{fig:sigma_vs_mag}
\end{figure}

Before analysing the photometric redshift properties of the radio source population specifically, it is useful to first verify the overall redshift accuracy of the estimates in the respective fields.
For galaxy evolution studies (where the overall bias is less critical), the two most important metrics are typically the robust scatter, $\sigma_{\textup{NMAD}}$, and outlier fraction, $O_{f}$.
Figures~\ref{fig:sigma_vs_z} and ~\ref{fig:sigma_vs_mag} illustrate how these metrics vary with redshift and magnitude for the full Bo\"{o}tes and COSMOS spectroscopic samples. In Table~\ref{tab:photzstats} we also present the corresponding photo-z quality metrics for the full spectroscopic sample and all subsets of radio detected sources.

\begin{table*}
	\centering
	\caption{Definitions of statistical metrics used to evaluate photometric redshift accuracy and quality along with notation used throughout the text.}\label{tab:definitions}

	\begin{tabular}{rll} 
		\hline
		Metric &  & Definition \\
		\hline
		$\sigma_{f}$ & Scatter - all galaxies & $ \text{rms}(\Delta z / (1+z_{\text{spec}}))$ \\
		$\sigma_{\text{NMAD}}$ & Normalised median absolute deviation & $1.48 \times \text{median} ( \left | \Delta z \right | / (1+z_{\text{spec}}))$ \\
		Bias & &$\text{median} (\Delta z )$\\
		O$_{f}$ & Outlier fraction & Outliers defined as $\left | \Delta z \right | / (1+z_{\text{spec}}) > 0.2$ \\
		$\sigma_{\text{O}_{f}}$  & Scatter excluding O$_{f}$ outliers & $ \text{rms}[\Delta z / (1+z_{\text{spec}})]$ \\
		$\overline{\textup{CRPS}}$ & Mean continuous ranked probability score &  $\overline{\textup{CRPS}} = \frac{1}{N} \sum_{i=1}^{N} \int_{-\infty}^{+\infty} [ \textup{CDF}_{i}(z) -  \textup{CDF}_{z_{s},i}(z)]^{2} dz$ - 
		\citet{2001math......1268R} \\
		\hline
	\end{tabular}
\end{table*}

\begin{table*}
\centering
\caption{Photometric redshift quality statistics for the Bo\"{o}tes (left) and COSMOS (right) spectroscopic samples. The statistical metrics (see Table~\ref{tab:definitions}) are shown for the full spectroscopic sample, the radio detected sources and for various subsets of the radio population. For each subset, values from the best performing are highlighted in bold font.}\label{tab:photzstats}
\begin{tabular}{c|ccccccccccccc}
\hline
& \multicolumn{6}{c}{Bo\"{o}tes} & & \multicolumn{6}{c}{COSMOS}\\
\hline
Templates & $\sigma_{f}$ & $\sigma_{\text{NMAD}}$ & Bias & O$_{f}$ & $\sigma_{\text{O}_{f}}$ & $\overline{\textup{CRPS}}$ 
& & $\sigma_{f}$ & $\sigma_{\text{NMAD}}$ & Bias & O$_{f}$ & $\sigma_{\text{O}_{f}}$ & $\overline{\textup{CRPS}}$ \\
\hline
\multicolumn{7}{c}{All Sources} & 
\multicolumn{7}{c}{All Sources}\\
EAZY & 0.772 & 0.040 & 0.007 & 0.111 & 0.049 & 0.293 
 &  & \textbf{0.365} & \textbf{0.019} & \textbf{-0.002} & \textbf{0.096} & \textbf{0.048}& \textbf{0.157} \\
Atlas &  0.879 & \textbf{0.037} & \textbf{-0.005} & 0.128 & \textbf{0.046} & 0.314 
 &  & 0.460 & 0.025 & -0.004 & 0.103 & 0.051 & 0.174 \\
XMM-C & \textbf{0.658} & 0.055 & -0.010 & \textbf{0.095} & 0.060 & \textbf{0.218}
 &  & 0.384 & 0.070 & -0.029 & 0.214 & 0.071 & 0.249\\
\hline
\multicolumn{7}{c}{All Radio Sources} &
 \multicolumn{7}{c}{All Radio Sources}\\
EAZY & 0.573 & 0.038 & 0.010 & \textbf{0.120} & \textbf{0.044} & 0.274 
& & \textbf{0.241} & \textbf{0.016} & \textbf{-0.001} & \textbf{0.081} & \textbf{0.042} & \textbf{0.115} \\
Atlas & \textbf{0.471} & \textbf{0.037} & -0.005 & 0.144 & 0.046 & 0.257
 & & 0.241 & 0.020 & -0.001 & 0.088 & 0.045 & 0.121 \\
XMM-C & 0.566 & 0.056 & \textbf{-0.003} & 0.124 & 0.059 & \textbf{0.210}
 & & 0.236 & 0.038 & -0.013 & 0.104 & 0.054 & 0.133 \\
\hline
\multicolumn{7}{c}{Radio Sources - Non X-ray/IR/Opt AGN} &
 \multicolumn{7}{c}{Radio Sources - Non X-ray/IR/Opt AGN}\\
EAZY & 0.490 & 0.029 & 0.005 & 0.030 & \textbf{0.041} & 0.203
& & \textbf{0.365} & \textbf{0.019} & \textbf{-0.002} & \textbf{0.096} & \textbf{0.048} & \textbf{0.157} \\
Atlas & \textbf{0.407} & \textbf{0.027} & -0.003 & \textbf{0.027} & 0.042 & \textbf{0.132}
 & & 0.460 & 0.025 & -0.004 & 0.103 & 0.051 & 0.174\\
XMM-C & 0.551 & 0.047 & \textbf{-0.002} & 0.049 & 0.054 & 0.151
& & 0.384 & 0.070 & -0.029 & 0.214 & 0.071 & 0.249 \\
\hline
\multicolumn{7}{c}{Radio Sources - X-ray Detected} &
 \multicolumn{7}{c}{Radio Sources - X-ray Detected}\\
EAZY & 0.769 & 0.648 & 0.307 & 0.606 & \textbf{0.055} & 0.734
 & & 0.296 & \textbf{0.028} & 0.006 & \textbf{0.169} & \textbf{0.053} & \textbf{0.225}\\
Atlas &  0.848 & 0.594 & -0.064 & 0.681 & 0.067 & 0.858 
& & 0.283 & 0.052 & \textbf{0.001} & 0.175 & 0.062 & 0.240 \\
XMM-C & \textbf{0.569} & \textbf{0.261} & \textbf{0.010} & \textbf{0.436} & 0.089 & \textbf{0.487}
 & & \textbf{0.273} & 0.056 & -0.009 & 0.186 & 0.063 & 0.231\\
\hline
\multicolumn{7}{c}{Radio Sources - IR AGN} &
 \multicolumn{7}{c}{Radio Sources - IR AGN}\\
EAZY & 0.742 & 0.503 & 0.236 & 0.609 & \textbf{0.066} & 0.702 
& & 0.516 & \textbf{0.079} & 0.016 & \textbf{0.303} & \textbf{0.064} & 0.445  \\
Atlas & 0.751 & 0.546 & -0.298 & 0.775 & 0.083 & 0.965 
&  & \textbf{0.454} & 0.132 & \textbf{-0.005} & 0.316 & 0.078 & 0.433 \\
XMM-C & \textbf{0.695} & \textbf{0.288} & \textbf{-0.012} & \textbf{0.510} & 0.097 & \textbf{0.505} 
& & 0.513 & 0.142 & -0.036 & 0.393 & 0.080 & \textbf{0.406}\\
\hline
\multicolumn{7}{c}{Radio Sources - Opt AGN} &
 \multicolumn{7}{c}{Radio Sources - Opt AGN}\\
EAZY & 0.661 & 0.406 & 0.131 & 0.545 & \textbf{0.069} & 0.646 
& & \textbf{0.162} & \textbf{0.041} & \textbf{-0.002} & \textbf{0.186} & \textbf{0.056} & \textbf{0.248} \\
Atlas & 0.770 & 0.603 & -0.317 & 0.769 & 0.070 & 1.046 
& & 0.175 & 0.064 & -0.014 & 0.211 & 0.060 & 0.292 \\
XMM-C & \textbf{0.500} & \textbf{0.187} & \textbf{0.000} & \textbf{0.430} & 0.077 & \textbf{0.503} 
& & 0.214 & 0.055 & -0.020 & 0.233 & 0.064 & 0.364  \\
\hline
\multicolumn{7}{c}{Radio Sources - $\log_{10}(L_{150\textup{MHz}} \textup{[W / Hz]}) > 25$}  &
\multicolumn{7}{c}{Radio Sources - $\log_{10}(L_{150\textup{MHz}} \textup{[W / Hz]}) > 25$} \\
EAZY & \textbf{0.477} & 0.081 & \textbf{0.014} & 0.341 & \textbf{0.046} & 0.478 & 
& \textbf{0.218} & \textbf{0.015} & \textbf{0.000} & \textbf{0.094} & \textbf{0.041} & \textbf{0.123} \\
Atlas & 0.539 & 0.290 & -0.049 & 0.512 & 0.052 & 0.765 
& & 0.233 & 0.021 & -0.001 & 0.099 & 0.044 & 0.135  \\
XMM-C & 0.498 & \textbf{0.074} & -0.033 & \textbf{0.279} & 0.075 & \textbf{0.403} 
& &  0.225 & 0.036 & -0.011 & 0.104 & 0.053 & 0.137 \\
\end{tabular}
\end{table*}

As expected given the availability of medium band observations, the COSMOS photo-zs (Fig.~\ref{fig:sigma_vs_z}) typically have lower scatter than the Bo\"{o}tes dataset at any given redshift.
However, at $z \lesssim 1$ the photo-zs for all template sets perform well in both fields, with $0.03 \lesssim \sigma_{\textup{NMAD}} \lesssim 0.05$ (Bo\"{o}tes) and $0.01 \lesssim \sigma_{\textup{NMAD}} \lesssim 0.03$ (COSMOS).

For both samples we find that the redshift estimates for X-ray detected and Opt/IR AGN population (dashed lines) typically perform worse on average than the remaining galaxy population at $z < 2$.
However, at $z\gtrsim2.5$ the two populations begin to converge to equivalent levels of scatter and outlier fraction.
While the `normal' galaxies at $z>2$ deteriorate in quality (likely a result of decreasing S/N - Figure~\ref{fig:sigma_vs_mag}), the photo-z estimates for sources in the X-ray/Opt/IR AGN sample begin to improve.
This convergence at higher redshift is potentially driven by the increasing importance of the common Lyman break feature in determining the fitted redshift.

While the primary goal of this paper is to draw conclusions on the relative photo-z accuracies for different source populations, it is also useful to compare the absolute accuracy of the photo-zs produced relative to those of high-quality sets available in the literature.
Therefore, in addition to the comparison between template sets, in Fig.~\ref{fig:sigma_vs_z} we also present the quality metrics of the published COSMOS2015 photometric redshift set \citep{Laigle:2016ku} for the same spectroscopic sample (the catalog `photoz' column; grey lines).
Because the COSMOS2015 estimates are not optimised for AGN (and exclude estimates for some X-ray sources), photo-zs for X-ray detected galaxies are taken from the results of \citet{Marchesi:2016eo}.
For the `normal' galaxy population, the scatter and outlier fractions of the EAZY and Atlas template sets perform comparably to the official COSMOS2015 estimates.
 In contrast, for the \citet{Laigle:2016ku} estimates alone the X-ray/IR/opt AGN sample perform significantly worse than the best estimates from this analysis.
Incorporating the photo-zs for X-ray sources from \citet[][as shown in Fig.~\ref{fig:sigma_vs_z}]{Marchesi:2016eo} the combined literature photo-zs performance improves, with scatter and outlier fractions at $z < 2$ comparable to  the best estimates from this analysis, but a poorer performance above this range.

For both the Bo\"{o}tes and COSMOS samples we find that the EAZY and Atlas template sets perform comparably and typically produce the lowest scatter ($\sigma_{\textup{NMAD}}$) in both the full spectroscopic sample and the full radio selected population.
However, in the sub-samples of X-ray detected sources or IR AGN, we find no consistency between the two different datasets.
In the wide area dataset, the XMM-COSMOS template set performs significantly better in almost all metrics than the other two sets, for AGN populations (see Table~\ref{tab:photzstats} .
Conversely, in the deep field the XMM-COSMOS set performs worst for the key $\sigma_{\textup{NMAD}}$ and $O_{\textup{f}}$ metrics in the subset of X-ray/Opt/IR AGN sources.

Given the consistent methodology used for both datasets, the underlying reason for this discrepancy is likely due to the differences in the source populations included in the relevant spectroscopic samples (see Section~\ref{sec:mw_agn}).
As seen in Fig.~\ref{fig:mag_vs_z}, the Bo\"{o}tes X-ray/Opt/IR AGN source population is typically significantly optically brighter than that probed in COSMOS and may therefore have intrinsically different SEDs.

One clear conclusion that can be drawn from Figures~\ref{fig:sigma_vs_z}/\ref{fig:sigma_vs_mag} and Table~\ref{tab:photzstats} is that there is no single template set which performs consistently best across all subsets and datasets.
Differences in the redshifts estimated by the three different template sets are found to systematically depend strongly both on optical magnitude (a proxy for overall S/N) and redshift.
Specifically, as sources become optically fainter the range between the highest and lowest predicted redshifts systematically increases.
As a function of redshift, this range of predicted photo-zs also increases significantly between $1 \lesssim z \lesssim 3$; above $z\sim3$ the estimates begin to converge again.
We see these trends in both the wide and deep fields, leading us to conclude that the redshift effect is not due to the systematics of the available optical data itself (e.g. the relatively shallow near-IR data in Bo\"{o}tes).

\subsection{Relative photo-z accuracy for radio and non-radio sources}\label{sec:radio_vs_nonradio}

It is clear that the absolute values for photometric redshift quality metrics are strong functions of the redshifts being probed, along with relative depth (S/N), resolution and wavelength coverage of the photometry available. The fundamental question for photometric redshift estimates in deep radio continuum surveys is how does the redshift accuracy differ between the radio detected and radio undetected source populations?

To understand how the different intrinsic source populations affect the resulting photo-z accuracy, we therefore measured the relative photo-z scatter and outlier fraction between the radio detected and non detected populations as a function of redshift.
To minimise the effects of known biases or photo-z quality dependencies, we first carefully match the two samples in redshift, magnitude and colour space.

Within the 3 dimensional parameter space of the spectroscopic redshift, $I$-band ($i^{+}$) magnitude and $I - 3.6\mu m$ ($i^{+} - 3.6\mu m$) colour, we calculate the 10 nearest neighbours for each radio detected source.
Due to the limited number of spectroscopic redshifts available, sources in the non-radio sample are allowed to be in the matched sample for more than one radio source.
Next, for each redshift bin, we calculate $\sigma_{\textup{NMAD}}$ and $O_{f}$ for the two matched samples and use a simple bootstrap analysis to estimate the corresponding uncertainties in these metrics.

\begin{figure}
\centering
	\includegraphics[width=0.99\columnwidth]{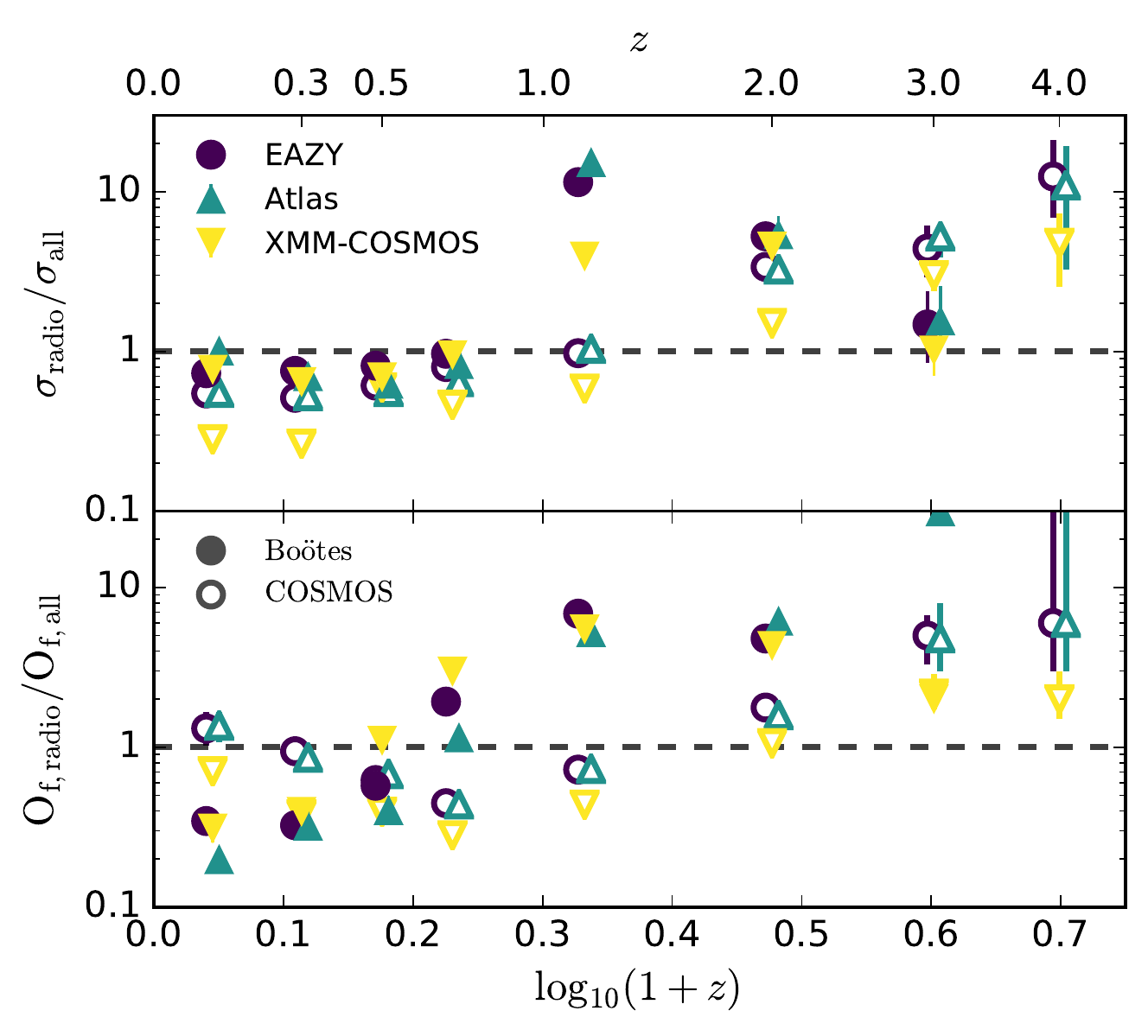}
  \caption{Photometric redshift quality for radio detected sources $\sigma_{\textup{radio}}$ ($O_{\textup{f, radio}}$) relative to matched samples of sources with no radio detection $\sigma_{\textup{all}}$  ($O_{\textup{f, all}}$) as a function of redshift, where $\sigma$ and $O_{\textup{f}}$ correspond to the normalised median absolute deviation and the outlier fraction as defined in Table~\ref{tab:definitions}. Details of sample matching procedure are outlined in Section~\ref{sec:radio_vs_nonradio}. In each plot we show the values for the EAZY (circles), Atlas (upward triangles) and XMM-COSMOS (downward triangles) template estimate for both the Bo\"{o}tes field (filled symbols) and COSMOS fields (empty symbols).}
  \label{fig:sigmaratio_vs_z}
\end{figure}

In Fig.~\ref{fig:sigmaratio_vs_z} we show the relative scatter and outlier fractions of these two matched samples.
We find that up to $z\sim1$ (where both spectroscopic samples are fairly representative), photometric redshifts estimated for radio sources have typically lower scatter and outlier fraction than galaxies with no radio detection that have similar magnitudes.
This trend is true for both datasets and for all three template sets.

Above $z\sim1$, photo-zs for radio sources are significantly worse than their matched non-radio detected counterparts.
This trend of increasing scatter/outlier fraction with redshift is not unexpected given as redshift increases the radio detected sources are increasingly luminous AGN for which photo-z estimates are expected to struggle.

In the Bo\"{o}tes field specifically (filled symbols), we see that at $z\sim1$ there is a significant jump in the measured scatter for radio sources.
Inspecting the magnitude-redshift distribution of the radio sample reveals that $z\sim1$ ($\log_{10}(1+z) \approx 0.3$) marks the transition where the AGES spectroscopic sample become dominated by the AGN selection criteria and almost all sources are classified as either X-ray sources or IR AGN.
We note however that the sample bias towards X-ray and IR AGN sources is true for both the radio and matched non radio samples, indicating that at higher redshift the radio-loud subset of X-ray/IR AGN sources is systematically more difficult to fit than the radio-faint population of similar magnitude.

\subsection{Photometric redshift accuracy as a function of radio power}\label{sec:radio_power}
In Section~\ref{sec:radio_vs_nonradio} we saw that the photo-zs for the radio detected population becomes systematically worse at high redshift.
If this trend is driven by the evolution in sample radio luminosities from the flux limited samples, we expect to observe the same trend when looking at a fixed redshift but evolving radio luminosity.
\begin{figure}
\centering
	\includegraphics[width=0.99\columnwidth]{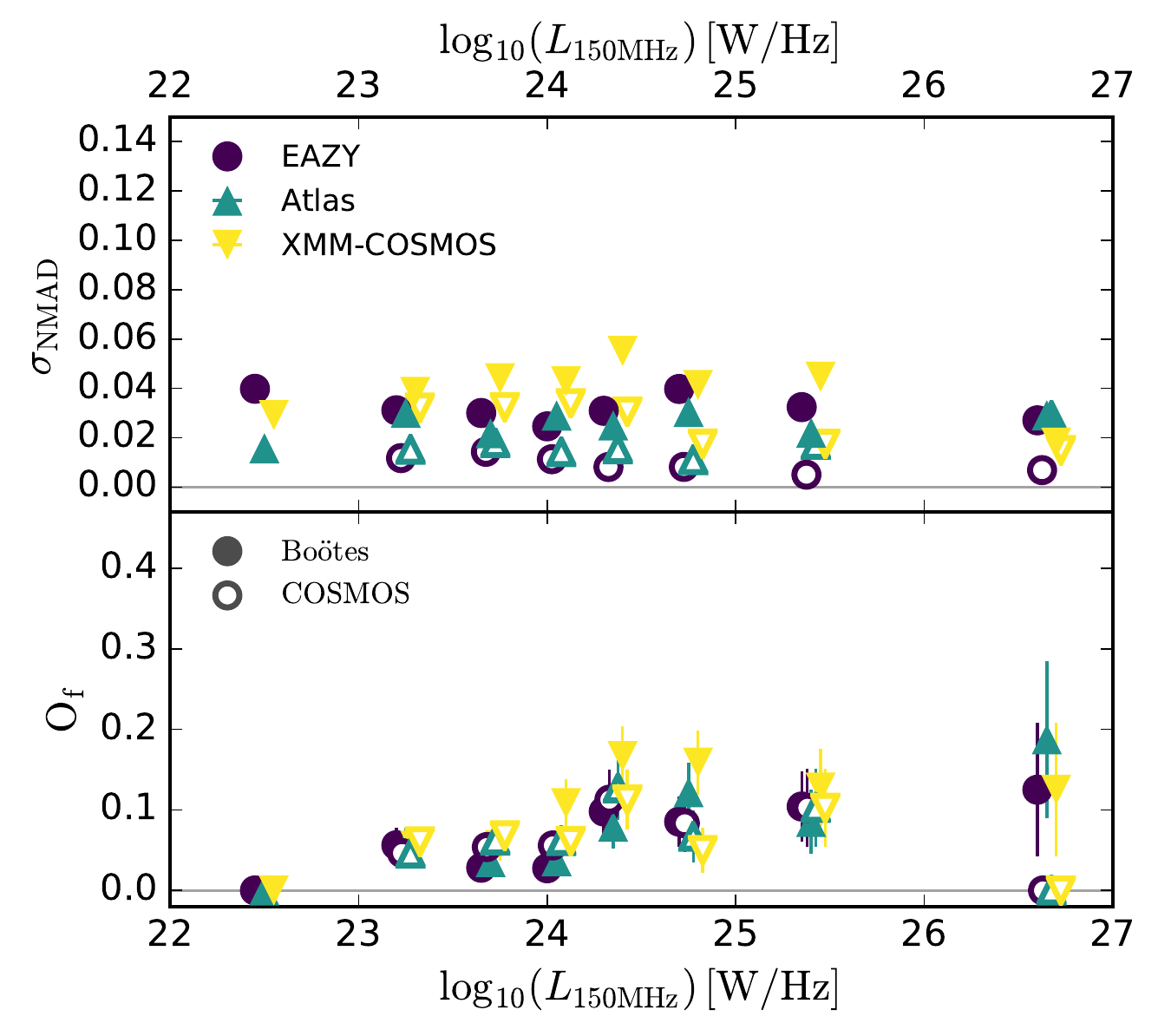}
	\includegraphics[width=0.99\columnwidth]{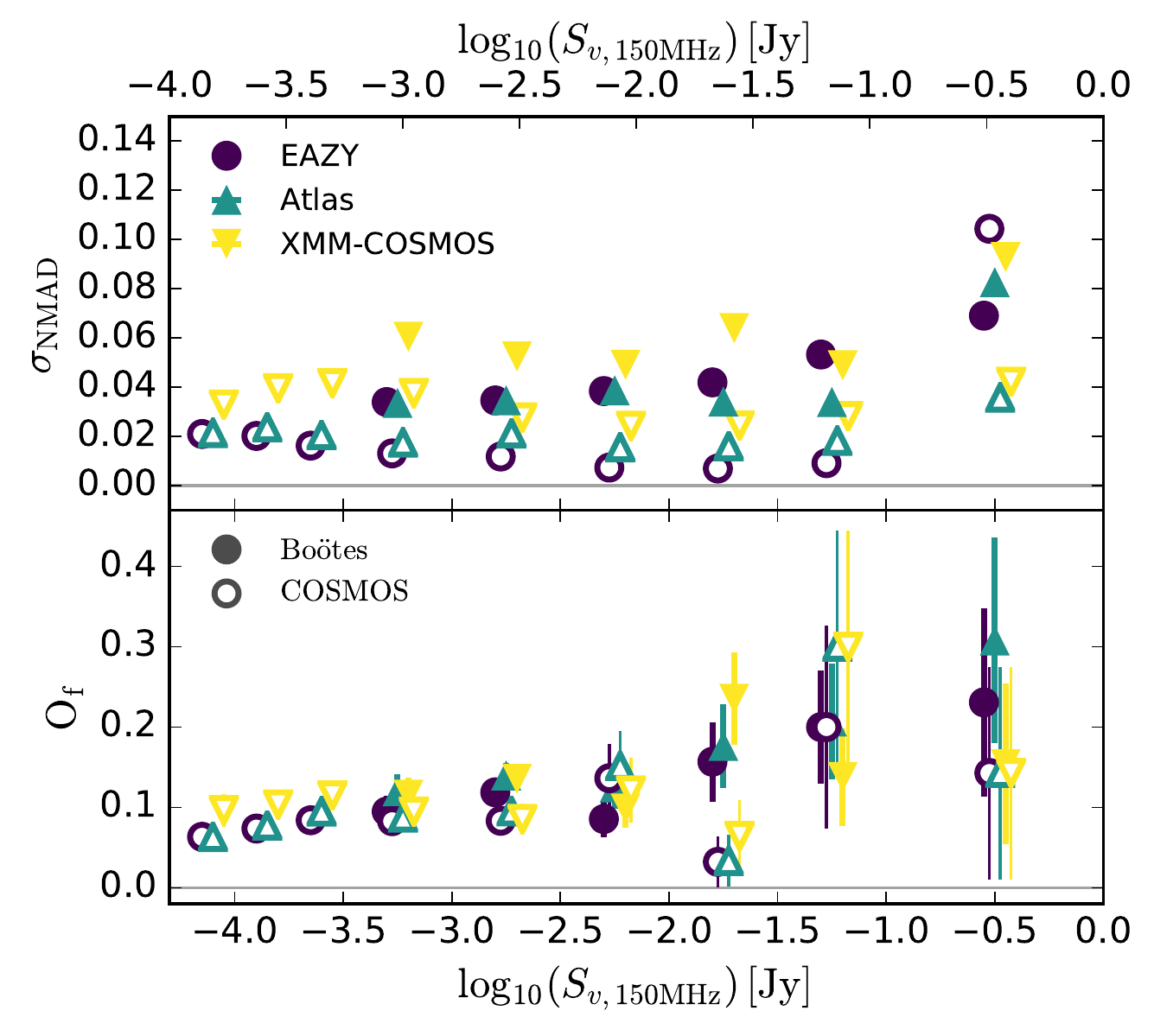}
  \caption{Photometric redshift scatter ($\sigma_{\textup{NMAD}}$; upper panels) and outlier fraction ($O_{f}$; lower panels) as a function of 150MHz radio luminosity (top) and flux (bottom) for galaxies within the redshift range $0.2 < z < 0.9$. In each plot we show the values for the EAZY (circles), Atlas (upward triangles) and XMM-COSMOS (downward triangles) template estimate for both the Bo\"{o}tes field (filled symbols) and COSMOS fields (empty symbols). Symbols have been offset horizontally only for clarity, all luminosity/flux bins are identical. Error-bars plotted for the outlier fractions illustrate the binomial uncertainties on each fraction.}
  \label{fig:sigma_radio_power}
\end{figure}
In Fig.~\ref{fig:sigma_radio_power} we present the evolution in $\sigma_{\textup{NMAD}}$ and $O_{f}$ as a function of $\log_{10}(L_{150\textup{MHz}})$ for sources with spectroscopic redshift in the range $0.2 < z < 0.9$.
We choose this redshift range because based on Fig.~\ref{fig:sigma_vs_z} we know that the scatter and outlier fraction of the full sample do not evolve strongly across this range.
In the COSMOS field, the scatter remains relatively constant with redshift for both the AGN and normal galaxy samples while the outlier fraction actually decreases slightly over this range.

In contrast to the naive expectation of increasing scatter with radio luminosity, it is evident that there is no clear evolution in $\sigma_{\textup{NMAD}}$ across the $\sim4$ orders of magnitude in radio luminosity probed by our samples.
The measured scatter follows a similar trend when examined as a function of radio flux.
Between $< 1$ mJy and 100 mJy, $\sigma_{\textup{NMAD}}$ remains effectively constant for all three template libraries in both and both datasets.
It is only at the very brightest radio fluxes ($0.1 < S_{\nu, 150\textup{MHz}} < 1$ Jy) where scatter increases for this redshift regime.
 
Some evidence of increasing outlier fraction as a function of $\log_{10}(L_{150\textup{MHz}})$ exists for both the deep and wide fields, with $O_{f}$ rising by $\sim 2$ between $\log_{10}(L_{150\textup{MHz}}) \approx 23$ and $\log_{10}(L_{150\textup{MHz}}) \gtrapprox 26$.
As a function of radio flux, the trend of increasing outlier fraction is even more pronounced.
Although there is significant scatter in the outlier fraction values and the small samples available at high radio power result in significant uncertainties, the trend is consistent across all three template sets.

In both fields, the overall AGN selected population has a higher outlier fraction across the redshift range ($0.2 < z < 0.9$) than the galaxy population. The rise in outlier fraction with radio flux illustrated in Fig.~\ref{fig:sigma_radio_power} may therefore be a result of the increasing radio AGN fraction with radio flux (/luminosity), see e.g. \citet{2013MNRAS.436.3759B}.

As radio surveys push to lower radio luminosities at higher redshift (e.g. $\log_{10}(L_{150\textup{MHz}}) < 25$ and $z\sim2$), these results suggest that photometric redshifts for the large samples of intermediate power AGN will be very comparable to those of `normal' galaxies.
However, what we cannot measure here is the effect of how the intrinsic SEDs of these sources might change over these redshifts and the resulting effects on photo-z estimates.

\subsubsection{Best-fitting templates vs radio power}

As radio luminosity increases and the source population becomes dominated by increasingly powerful AGN, a plausible expectation is that photo-z template sets incorporating AGN/QSO templates will perform better than stellar-only template sets.
Equally, at low radio luminosities where the population becomes dominated by star-forming galaxies, one would expect the template sets optimised for stellar emission to provide the most accurate photo-zs.
Based on the results of Fig.~\ref{fig:sigma_radio_power} however, there is no clear dependence of the preferred template on radio luminosity.

This result is in line with our expectation on the radio source population; namely its extremely diverse nature.
Across a broad range of radio luminosities, the observed spectral energy distributions are consistent with sources ranging from radio galaxies with old stellar populations to star-forming galaxies and luminous QSOs.

\subsection{Accuracy of the redshift PDFs}\label{sec:pdfcalibration}
While the scatter between the estimated photometric redshift (whether that is the peak or median of the $P(z)$) and the spectroscopic redshifts is a useful metric for judging their accuracy, this does not take into account the uncertainties on individual measurements nor the potential complexities of the $P(z)$ itself (e.g. multiple peaks or asymmetry).
In addition to ensuring the minimum scatter and outlier fraction possible, it is therefore essential that the estimated $P(z)$ accurately represent the true uncertainty of the photometric redshifts.
Even with the inclusion of additional photometry errors, it is common for template fitting photo-z codes to be over-confident in the predicted redshift accuracy \citep{Dahlen:2013eu,2016MNRAS.457.4005W}.

To quantify the over- or under-confidence of our photometric redshift estimates, we follow the example of \citet{2016MNRAS.457.4005W} and x where the spectroscopic redshift is just included.
For a set of redshift PDFs which perfectly represent the redshift uncertainty (e.g. 10\% of galaxies have the true redshift within the 10\% credible interval, 20\% within their 20\% credible interval, etc.), the expected distribution of $c$ values should be constant between 0 and 1.
The cumulative distribution, $\hat{F}(c)$, should therefore follow a straight 1:1 relation.
Curves which fall below this expected 1:1 relation therefore indicate that there is overconfidence in the photometric redshift errors; the $P(z)$s are too sharp.

\begin{figure}
\centering
	\includegraphics[width=0.5\columnwidth]{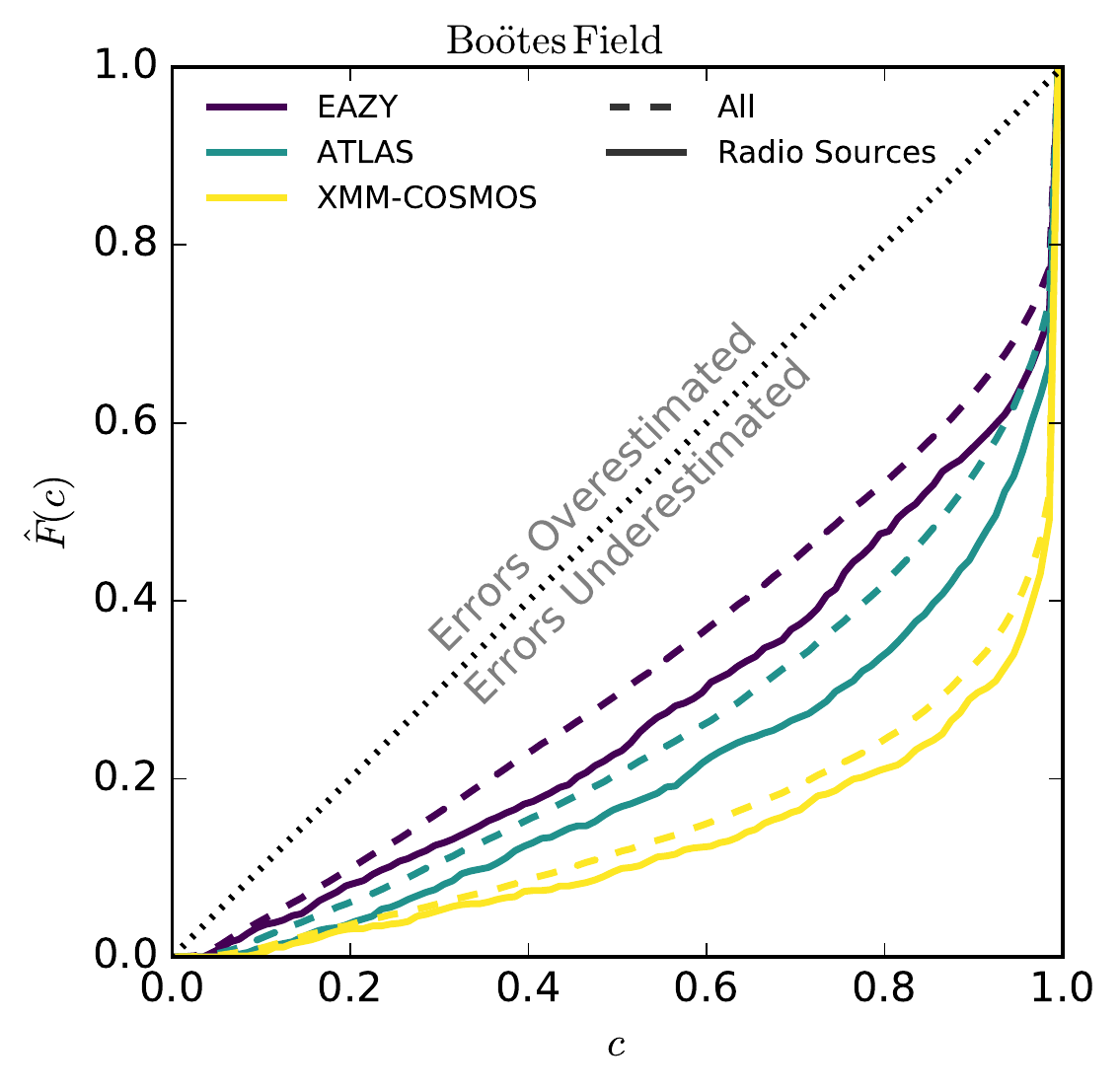}\includegraphics[width=0.5\columnwidth]{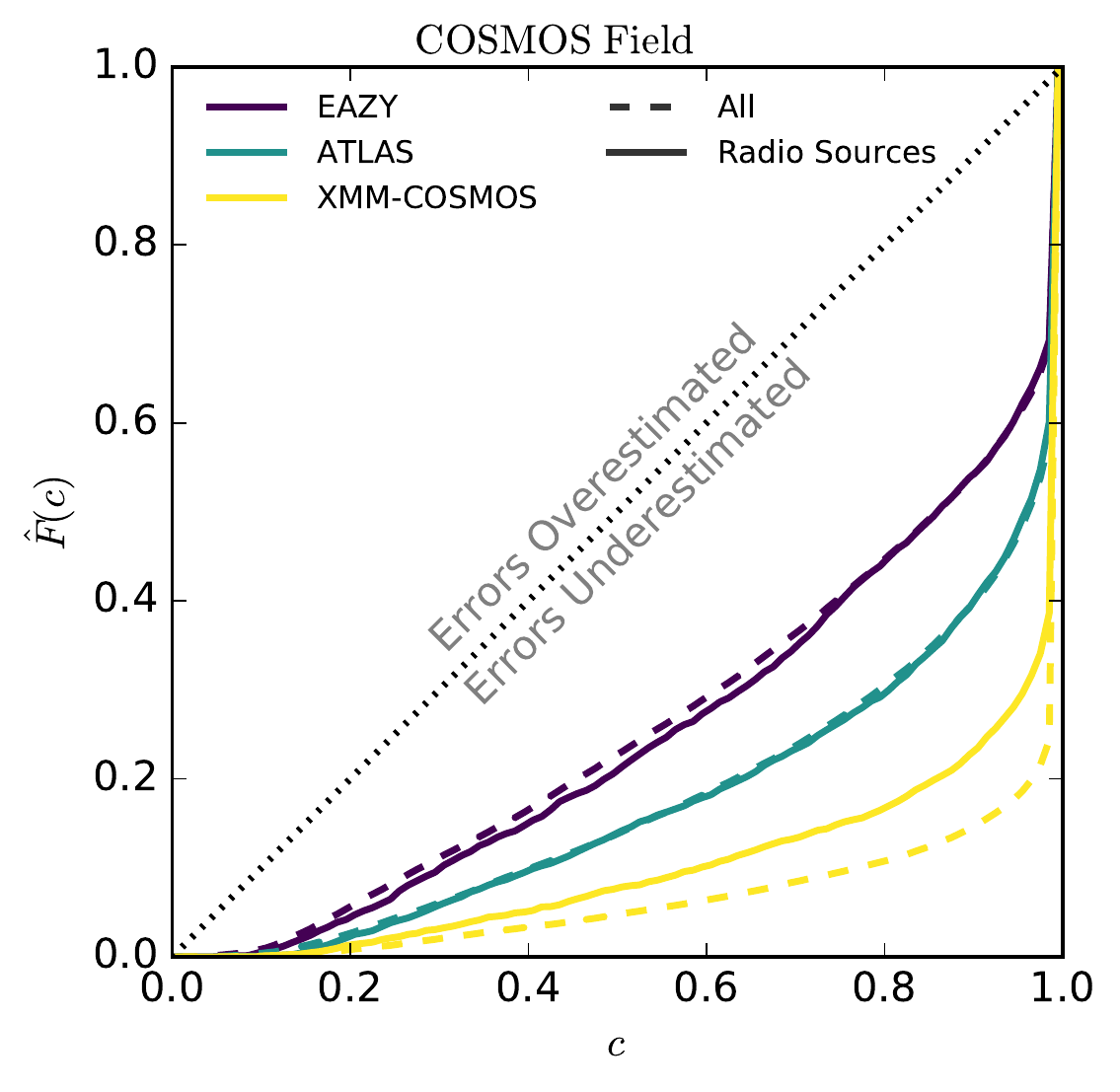}

  \caption{Q-Q ($\hat{F}(c)$) plots for the redshift PDFs for each template set, as directly produced by \textsc{eazy}. Plotted are the cumulative distributions for the full spectroscopic redshift sample (solid line) and for the sub-sample of LOFAR detected sources (dashed line).}
  \label{fig:qq_all}
\end{figure}

In Figure~\ref{fig:qq_all}, we show the $\hat{F}(c)$ distributions (Q-Q plots) for the uncorrected $P(z)$ output of each template set.
For both the full spectroscopic samples (dashed lines) and radio detected samples (solid lines), all three template sets show significant overconfidence in the photometric redshift errors.
The $P(z)$ estimates based on the EAZY template set are the most accurate while the XMM-COSMOS template set performs the worst.
We also find that despite having significantly lower scatter relative to the spectroscopic sample, the COSMOS field redshift estimates are noticeably more overconfident than those in the Bo\"{o}tes field.

Using a training subset of each population (AGN vs non-AGN), we smooth the redshift PDFs to minimise the euclidean distance between the observed $\hat{F}(c)$ and the desired 1:1 relation.
To do this we define the rescaled redshift PDF for a galaxy, $i$, as
\begin{equation}
	P(z)_{\textup{new}, i} \propto P(z)_{\textup{old}, i}^{1/\alpha(m_{i})},
\end{equation}\label{eq:smoothing_1}
where $\alpha(m)$ is a constant, $c$, below some characteristic apparent magnitude, $m_{c}$, and follows a simple linear relation above this magnitude, e.g.
\begin{equation}
	\alpha(m) = \begin{cases}
	 \alpha_{c} & m \leq m_{c}\\
	 \alpha_{c} + \kappa \times(m-m_{c}) & m > m_{c}.
	\end{cases}
\end{equation}\label{eq:smoothing_2}
For both datasets, we use the equivalent $I$/$i^{+}$ band optical magnitude for calculating the magnitude dependence.
We also assume a characteristic magnitude of $i^{+} = 20$ for the COSMOS sample \citep{Laigle:2016ku} and $I = 18$ for the shallower Bo\"{o}tes sample.
The parameters $c$ and $k$ are then fit using the \textsc{emcee} Markov Chain Monte Carlo fitting tool \citep[MCMC][]{ForemanMackey:2013io}.

\begin{figure*}
\centering
Bo\"{o}tes Field

	\includegraphics[width=0.33\textwidth]{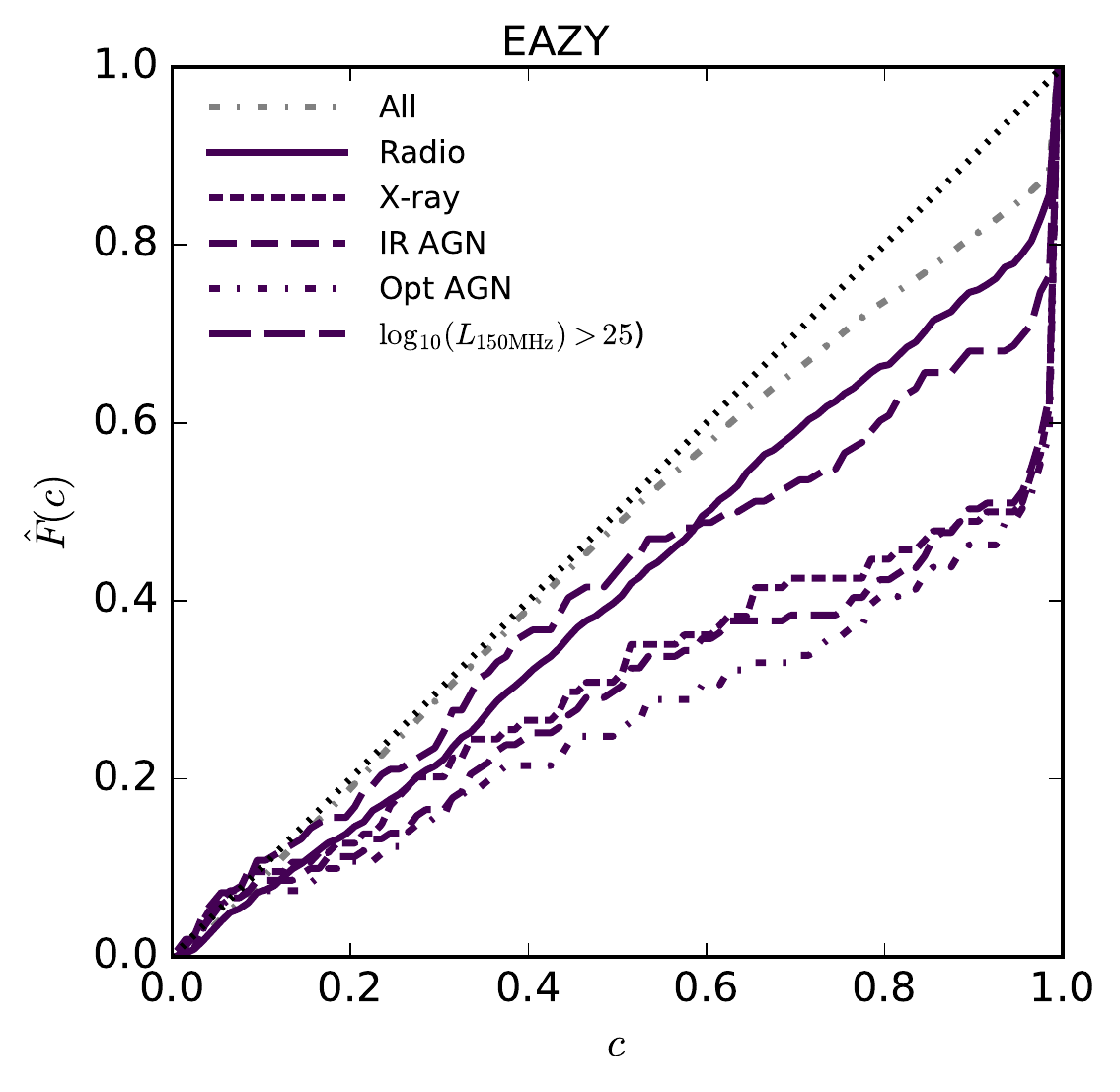}
    \includegraphics[width=0.33\textwidth]{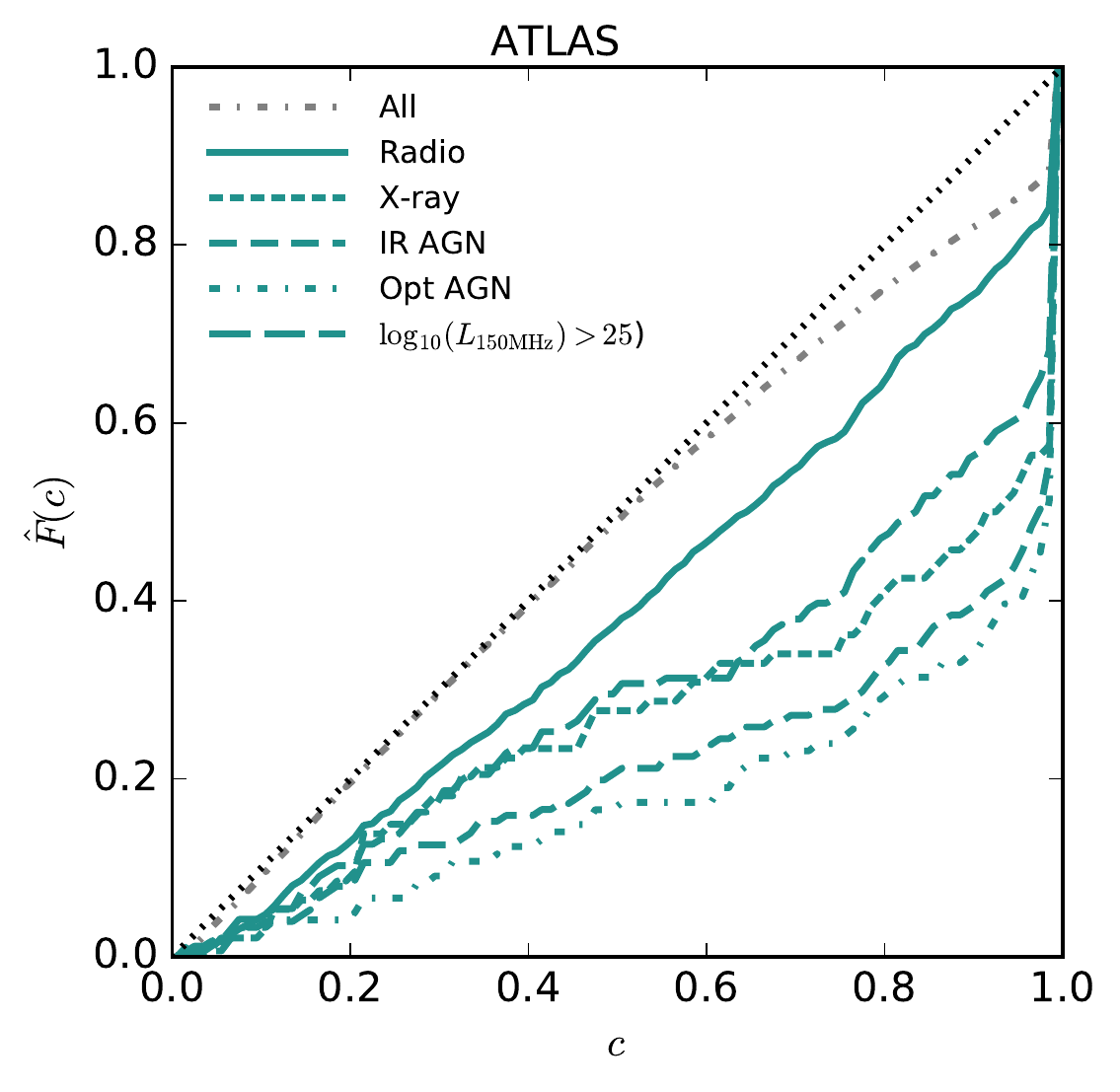}
	\includegraphics[width=0.33\textwidth]{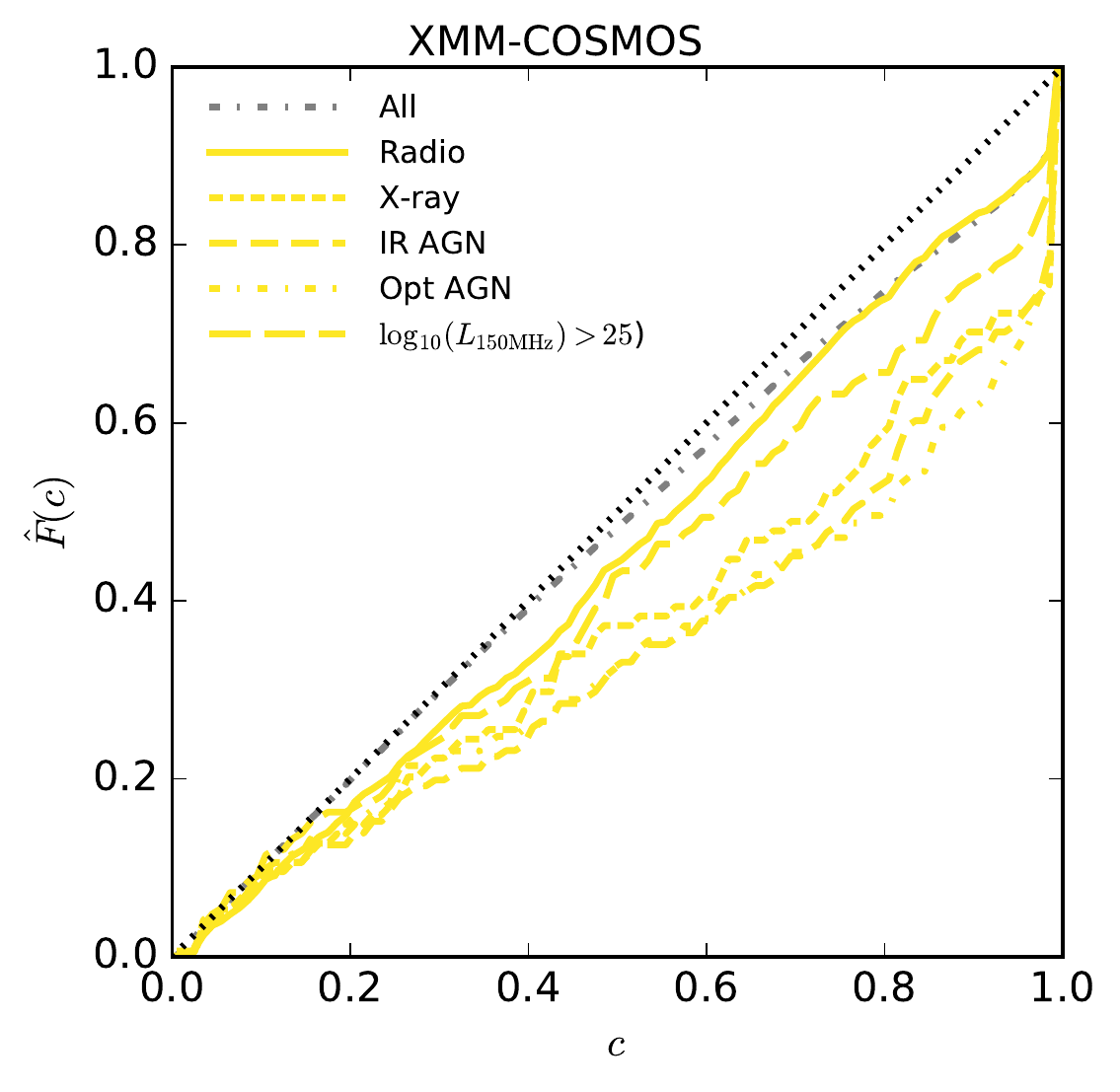}

  \caption{Q-Q ($\hat{F}(c)$) plots for the redshift PDFs for each template set in the Bo\"{o}tes field after smoothing the raw \textsc{eazy} PDFs to minimise the distance between the observed distribution and the ideal. Plotted are the cumulative distributions for all radio detected sources (solid lines) as well the subsets of the radio sample which are X-ray detected (short dashed lines), infrared AGN (medium dashed lines) and high power radio sources (long dashed lines).}
  \label{fig:qq_templates}
\end{figure*}

\begin{figure*}
\centering
COSMOS Field

	\includegraphics[width=0.33\textwidth]{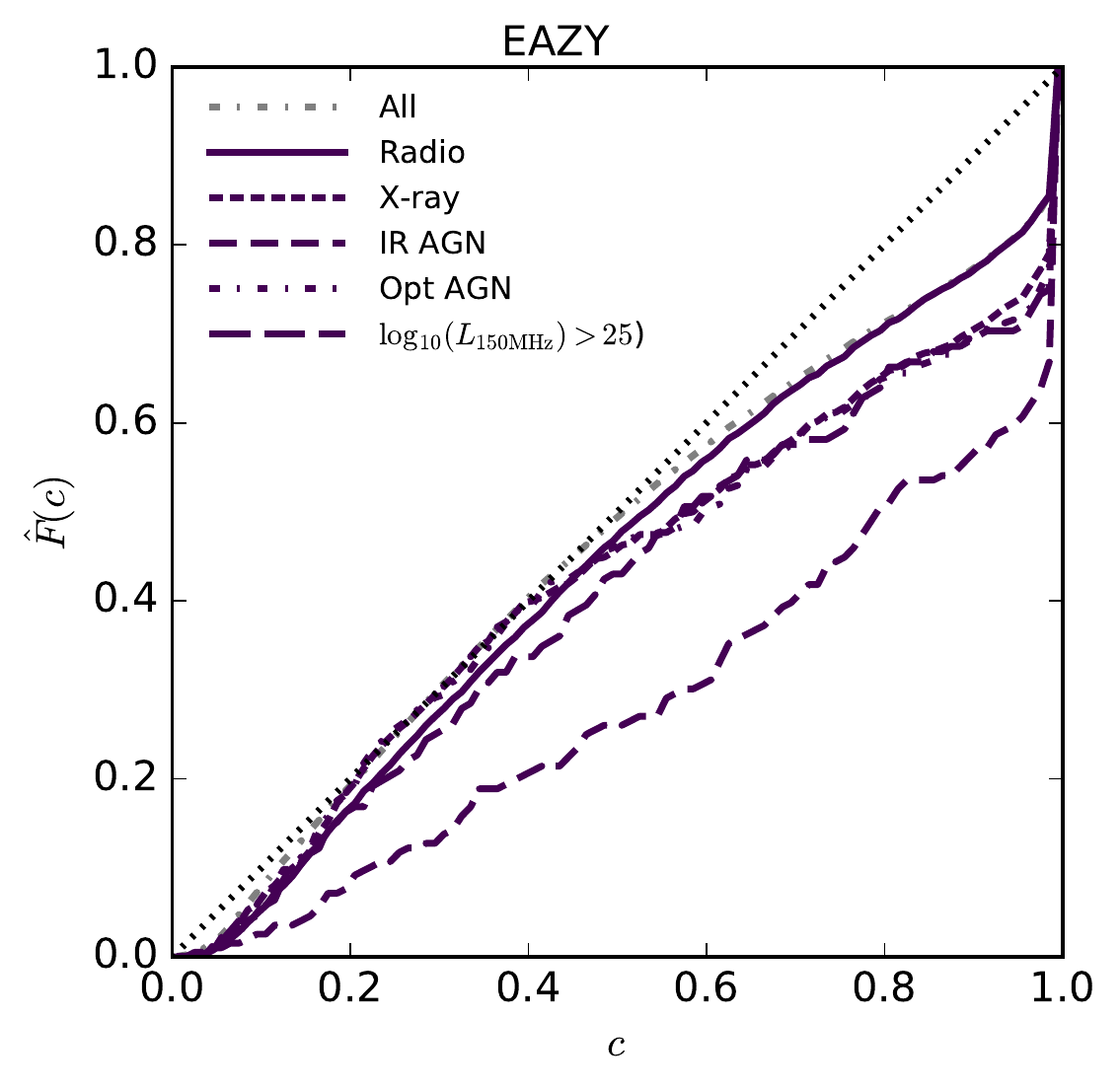}
    \includegraphics[width=0.33\textwidth]{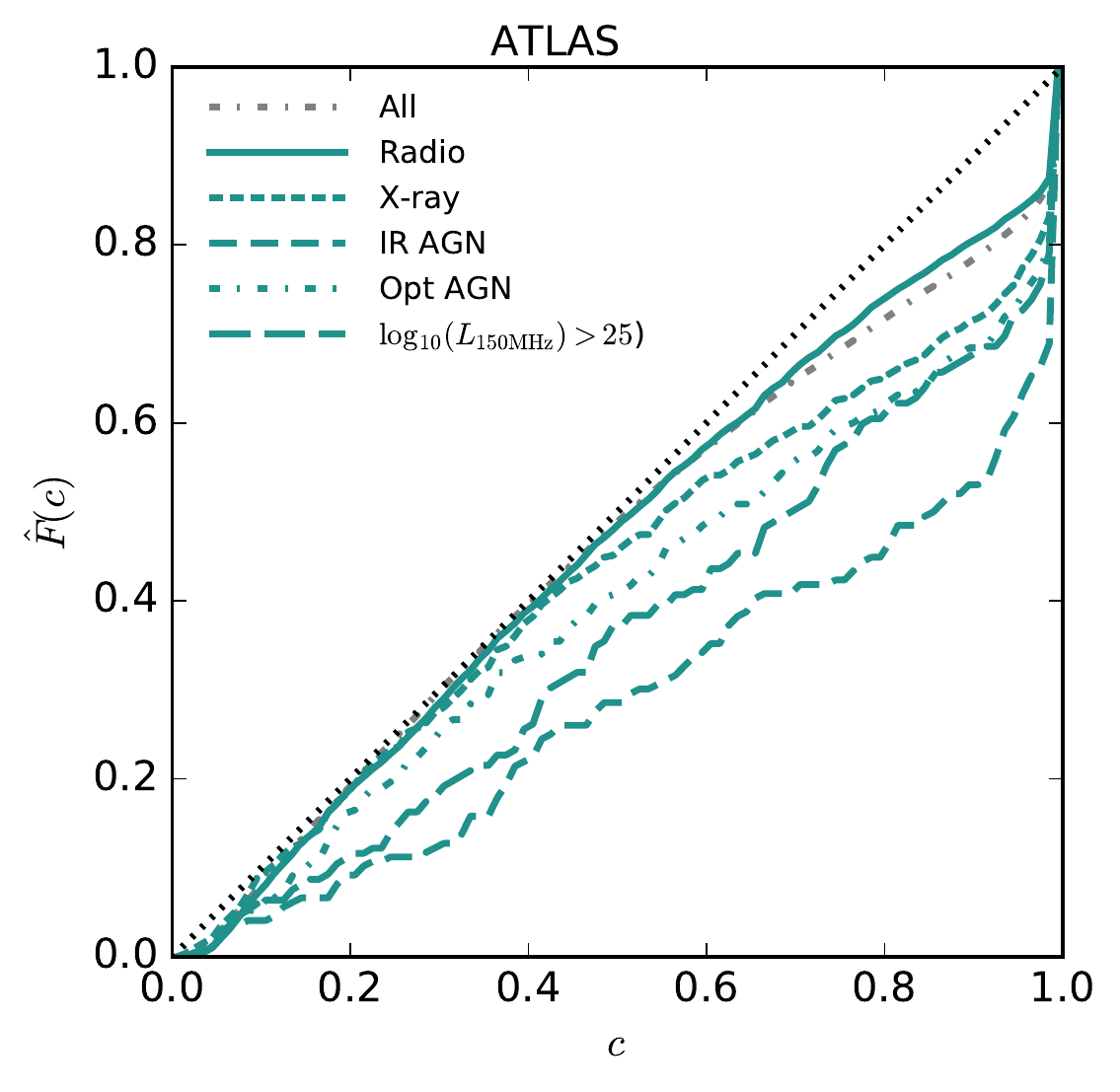}
	\includegraphics[width=0.33\textwidth]{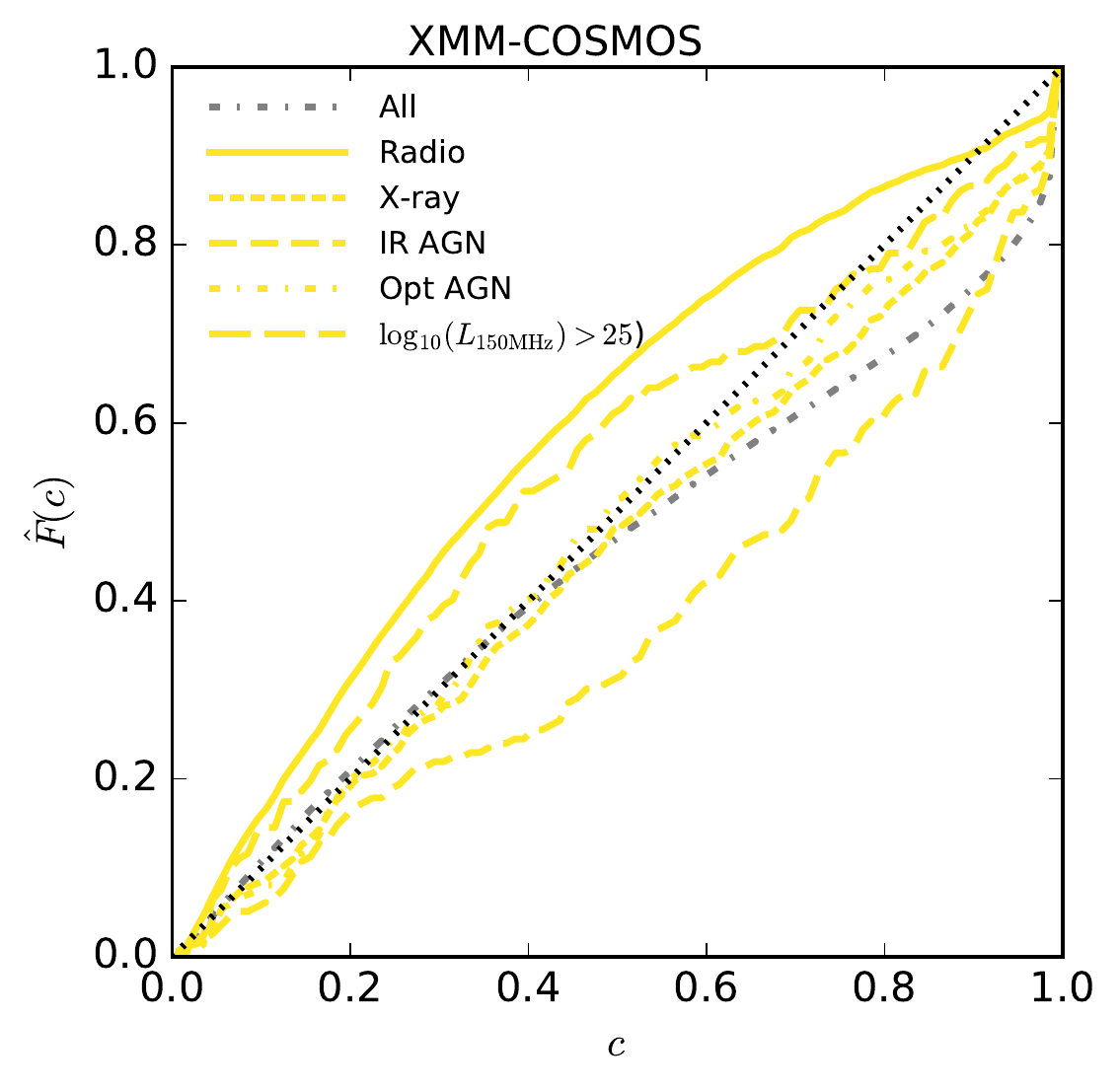}

  \caption{Q-Q ($\hat{F}(c)$) plots for the redshift PDFs for each template set in the COSMOS field after smoothing the raw \textsc{eazy} PDFs to minimise the distance between the observed distribution and the ideal. Lines as described in Fig.~\ref{fig:qq_templates}.}
  \label{fig:qq_templates_cosmos}
\end{figure*}

In Fig.~\ref{fig:qq_templates} and \ref{fig:qq_templates_cosmos} we show the resulting $\hat{F}(c)$ distributions for the all sources as well as the radio detected sources (and subsets thereof) after the redshift PDFs have been calibrated using the \emph{full spectroscopic redshift sample}.
For the Bo\"{o}tes field, the consensus PDFs from all three template sets are significantly improved for the full spectroscopic sample.
The 0 to 50\% credible interval ranges are all very accurately measured, with only a small remaining overconfidence within the 80\% credible interval.
For the COSMOS field, all three template sets perform well within the 50\% credible interval but the tails of the distributions are not as accurate as those for the Bo\"{o}tes field.

Although the calibrated redshift PDFs for the radio detected subsets are somewhat improved by the calibration procedure, they do not match the same accuracy as the wider spectroscopic redshift sample.
Of the three template sets, the calibrated PDFs from the XMM-COSMOS template set are the most accurate overall for the Bo\"{o}tes field.
However, for the COSMOS field, the calibrated PDFs for the XMM-COSMOS set are under-confident for the overall radio sample. 

For both the AGN and non-AGN calibration subsets the smoothing applied to the XMM-COSMOS set is significantly higher than required for the EAZY and Atlas template sets (with the exception of the Bo\"{o}tes AGN sample for the Atlas library; Table~\ref{tab:smoothing}).
The resulting PDFs (while accurately representing the uncertainty for the overall sample) are much broader than the other template sets.
The typical 1-2$\sigma$ uncertainties on an individual galaxy redshift solution are $\approx 2-3$ smaller for the EAZY/Atlas template sets compared to XMM-COSMOS.
 
\begin{table}
\centering	
\caption{Magnitude-dependent redshift PDF smoothing parameters derived during the error calibration procedure. The smoothing is applied following Eq.~\ref{eq:smoothing_1} and \ref{eq:smoothing_2} with the respective values of $\alpha_{c}$ and $\kappa$. Note, $m_{c} = 18$ and 20 for the Bo\"{o}tes and COSMOS fields respectively.}\label{tab:smoothing}
\begin{tabular}{cccccc}
\hline
& \multicolumn{2}{c}{Bo\"{o}tes} & & \multicolumn{2}{c}{COSMOS} \\
\hline
Templates & $\alpha_{c}$ &  $\kappa$ & & $\alpha_{c}$ &  $\kappa$ \\
\hline
\multicolumn{6}{c}{`Galaxies'} \\
EAZY & 1.50 & 0.41 & & 0.57 & 0.71 \\
Atlas & 1.04 & 1.29 & & 2.47 & 0.76 \\
XMM-COSMOS & 5.54 & 1.17 & & 6.36 & 4.73 \\
\hline
\multicolumn{6}{c}{`AGN'} \\
EAZY & 9.42 & 2.57 & & 3.21 & 0.50 \\
Atlas & 25.5 & 1.38 & & 3.00 & 1.59 \\
XMM-COSMOS & 10.7 & 2.22 & & 12.2 & 0.45 \\
\end{tabular}
\end{table}

\section{Optimized redshifts through hierarchical Bayesian combination}\label{sec:hbmethod}
As illustrated in Sections~\ref{sec:overall}-\ref{sec:radio_power}, no single template set can perform well for all types of radio-detected galaxy.
To obtain the best photometric redshift estimates for sources in future deep radio continuum surveys one would therefore ideally like to pre-classify every galaxy and fit it with the optimum method for that source type \citep[as successfully implemented by][]{Fotopoulou:2016fc}.
However, some of the key properties necessary for such \emph{a priori} classification are potentially not going to be known at the time photometric redshifts are fitted.

A potential solution to this problem lies in the combination of multiple photo-z within a Bayesian framework such as hierarchical Bayesian  (HB) combination \citep{Dahlen:2013eu,CarrascoKind:2014jg} or Bayesian model combination/averaging \citep{CarrascoKind:2014jg}.
Both of these ensemble methods for photometric redshifts have been illustrated to improve estimates for normal galaxy populations, with the combined redshift PDF more accurate and less biased than any individual photo-z determination incorporated in the analysis.

To further improve the photometric redshifts for radio continuum sources we therefore combine the estimates from each template set through a hierarchical Bayesian combination.

\subsection{Hierachical Bayesian combination of redshift PDFs}
Following the method outlined in \citet{Dahlen:2013eu}, a consensus $P(z)$ is determined while accounting for the possibility that individual \emph{measured} redshift probability distributions $P_{m}(z)_{i}$ are incorrect.
The possibility that an individual $P(z)$ is incorrect is introduced as a nuisance parameter, $f_{\text{bad}}$, that is subsequently marginalised over.

Following \citet{Dahlen:2013eu}, we define for each redshift estimate, $i$,
\begin{equation}
	\begin{split}
	P(z, f_{\text{bad}})_{i} = P(z|\text{bad measurement})_{i}f_{\text{bad}} \\
	+ P(z|\text{good measurement})_{i} (1-f_{\text{bad}}),
	\end{split}
\end{equation}
where $P(z|\text{bad measurement})$ ($U(z)$ hereafter for brevity) is the redshift probability distribution assumed in the case where the estimated $P_{m}(z)_{i}$ is incorrect and $P(z|\text{good measurement}) \equiv P_{m}(z)_{i}$ is the case where it is correct.
The choice of $U(z)$ is explored in detail in the following section.
For now, given a sensible choice of $U(z)$, the combined $P(z, f_{\text{bad}})$ for all $n$ measurements is then given by
\begin{equation}
	P(z, f_{\text{bad}}) = \prod_{i=1}^{n}P(z, f_{\text{bad}})_{i}^{1/\beta},
\end{equation}
where the additional hyper-parameter, $\beta$, is a constant that defines the degree of covariance between the different measurements.
For completely independent estimates $\beta = 1$, while for estimates that are fully covariant $\beta = n$ ($= 3$ in this work).
In this work we expect some reasonable degree of covariance between the three estimates as a result of the common photometric data and fitting algorithms used.
Although the peak of the final redshift distribution is independent of $\beta$, changes in $\beta$ do have an effect on the distribution widths.
As part of the hierarchical Bayesian combination, $\beta$ can also therefore be tuned such that posterior redshift distributions more accurately represent the redshift uncertainties.

Finally, we marginalise over $f_{\text{bad}}$ to produce the consensus redshift probability distribution for each object
\begin{equation}
	P(z) = \int^{f^{\text{max}}_{\text{bad}}}_{f^{\text{min}}_{\text{bad}}} P(z, f_{\text{bad}}) df_{\text{bad}},
\end{equation}
where $f^{\text{min}}_{\text{bad}}$ and $f^{\text{max}}_{\text{bad}}$ are the lower and upper limits on the fraction of bad measurements. While fixed by definition to lie in the range $0 \leq f_{\text{bad}} \leq 1$, the exact limits used when marginalising over $f_{\text{bad}}$ can also be tuned using the training sample \citep{CarrascoKind:2014jg}.

\subsubsection{Assumptions for the $U(z)$ prior}\label{sec:priors}
During the calculation of the consensus redshift PDF, it is necessary to make an assumption on what the redshift prior is in the case where a given measurement is bad.
The simplest assumption for $U(z)$ is that in cases where the measurement is bad, we have zero information on the redshift of a given object.
Therefore, $U(z)$ is a flat prior, whereby $U(z) = 1 / N $ for redshifts in the range of fitting $0 < z < N$.

As is discussed by \citeauthor{Dahlen:2013eu}, we can also assume a more informative prior such as one which is proportional to the redshift dependent differential comoving volume, $\mathrm{d} V(z)/\mathrm{d} z$.
Given the nature of the deep multi-wavelength surveys being used and the broad redshift range of interest, a volume prior increases the likelihood of sources being at higher redshifts and disfavours low-redshift solutions where $\mathrm{d} V(z)/\mathrm{d} z$ is very small.

Alternatively, as we adopt in our analysis, magnitude information for each source can also be incorporated through the use of an empirical or model-based magnitude prior \citep{Benitez:2000jr}.
The benefit of incorporating magnitude dependent redshift priors in template fitting has been well illustrated in the literature \citep{Benitez:2000jr,Hildebrandt:2012du,Molino:2014iz} and so the assumption of a magnitude dependent $U(z)$ is therefore well motivated.

Our empirical redshift prior is based on a modified version of the functional form outlined in \citet{Benitez:2000jr}.
Using subset of the spectroscopic training set, we fit the observed redshift - magnitude relation, $p(z|m_{I})$, with the function:
\begin{equation}\label{eq:prior_function}
  p(z|m_{I}) \propto \left ( c + \frac{ \mathrm{d} V(z)}{ \mathrm{d} z} \right ) \times \exp \left \{   -\left [   \frac{z}{z_{m}(m_{I})}    \right]^{\alpha}  \right \}.
\end{equation}
As in \citet{Benitez:2000jr}, the prior distribution at high redshifts is determined by an exponential cut-off above a magnitude-dependent redshift $z_{m}$.
However, rather than a linear dependence on $m_{I}$, we assume $z_{m} = z_{0} + k_{1}z + k_{2}z^{2}$.
Additionally, in place of the power-law term $z^{\alpha}$, we use the differential comoving volume element $\mathrm{d} V(z)/\mathrm{d} z$.
Following \citet{Hildebrandt:2012du}, we also include the additional parameter, $c$, to allow for non-vanishing likelihood as $z  \rightarrow 0$.
We make the assumption that for AGN selected sources, $c = 0$, while for all other sources $c = 0.001$.

\begin{figure}
 \centering
  \includegraphics[width=0.96\columnwidth]{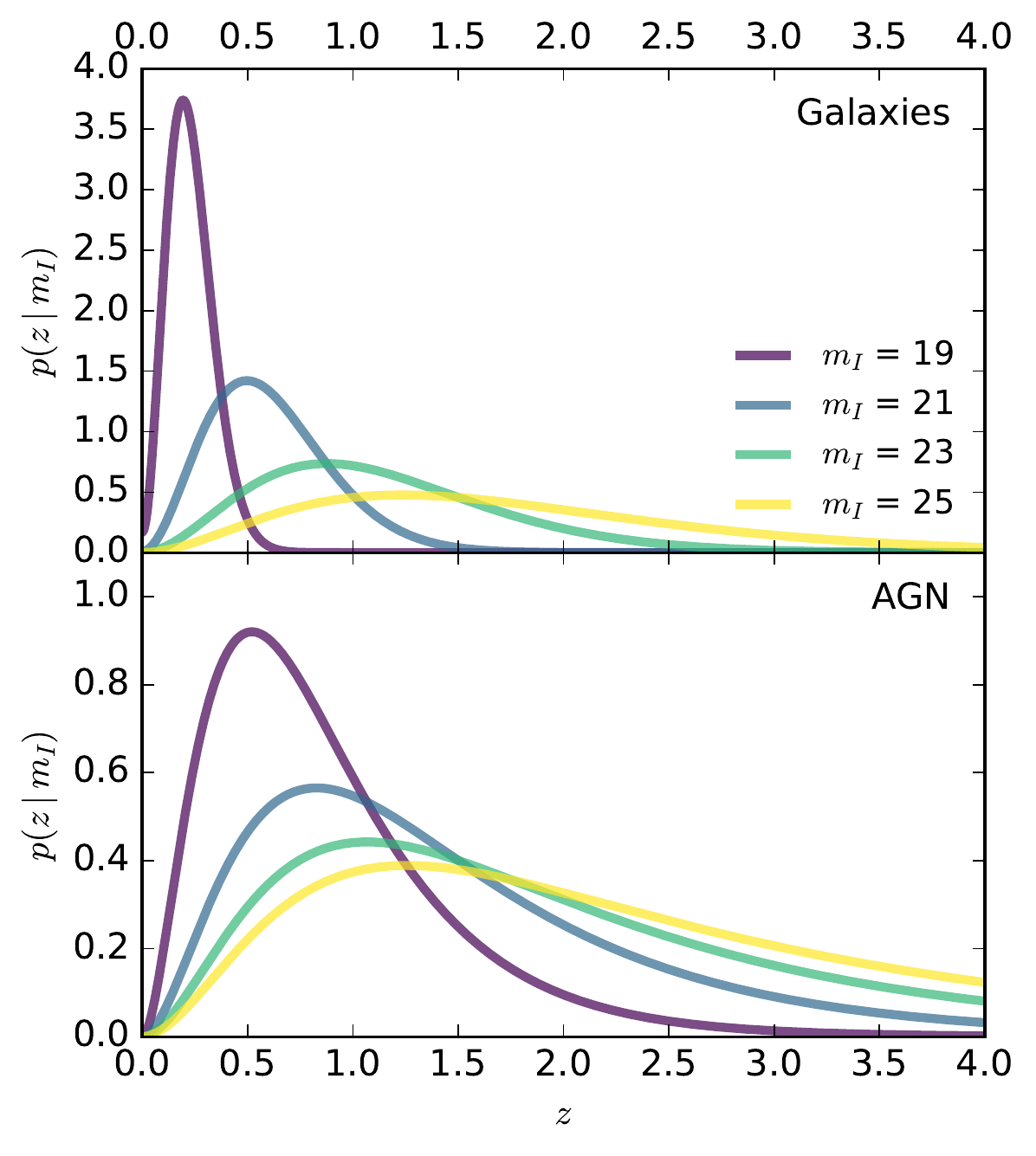}
  \caption{Empirical redshift priors for 4 different $I$-band magnitudes based on the functional form in Equation~\ref{eq:prior_function}. Top: redshift priors for sources which are not classified \emph{a priori} as AGN based on X-ray or IR AGN criteria. Bottom: redshift priors for sources which are classified as either X-ray sources or IR AGN (see Section~\ref{sec:mw_agn}).}
  \label{fig:priors}
\end{figure}

The parameters $\alpha$, $z_0$, $k_{1}$ and $k_{2}$ are estimated by fitting the functional form to the desired subset of test galaxy samples using MCMC \citep{ForemanMackey:2013io}.
Fig.~\ref{fig:priors} illustrates the resulting redshift priors as a function of $I$ magnitude for `normal' galaxies (top) and X-ray/IR/opt selected AGN (bottom).
At bright apparent optical magnitudes there is a clear difference in the redshift distribution between the two source populations, with AGN sources having both a higher median redshift and a more extended tail at higher redshifts ($z > 3$).

\subsubsection{Tuning of hyper-parameters using spectroscopic sample}
In addition to the assumption of $U(z)$, it is also necessary to assume or fit the additional hyper-parameters, $\beta$, $f^{\text{min}}_{\text{bad}}$ and $f^{\text{max}}_{\text{bad}}$.
Our assumptions for $f^{\text{min}}_{\text{bad}}$ and $f^{\text{max}}_{\text{bad}}$ are based on the measured and expected outlier fractions for the relevant source populations.
For non-AGN, we therefore marginalise $f_{\text{bad}}$ over the range $0 < f_{\text{bad}} < 0.05$ while for the X-ray/IR sample, we marginalise over the range $0 < f_{\text{bad}} < 0.5$.

As discussed in the previous section, $\beta$ can be tuned to maximise the accuracy of the resulting consensus $P(z)$ estimates.
We therefore fit $\beta$ using the spectroscopic training sample.
Specifically, we find the $\beta$ which minimises the distance between measured $\hat{F}(c)$ and the desired 1:1 relation within 80$\%$ HPD CI.
The restriction of fitting only within the 80$\%$ HPD CI region is motivated by the observation by \citet{2016MNRAS.457.4005W} that even for well calibrated photometric data, non-gaussian errors and uncalibrated template systematics can result in the tails of $P(z)$ distribution being poorly described.
By restricting the optimisation to only $<80\%$ HPD CI, we prevent over smoothing of the $P(z)$ due to these low probability tails.

\subsection{Optimised photometric redshift properties}
 \begin{figure}
 \centering
  \includegraphics[width=0.96\columnwidth]{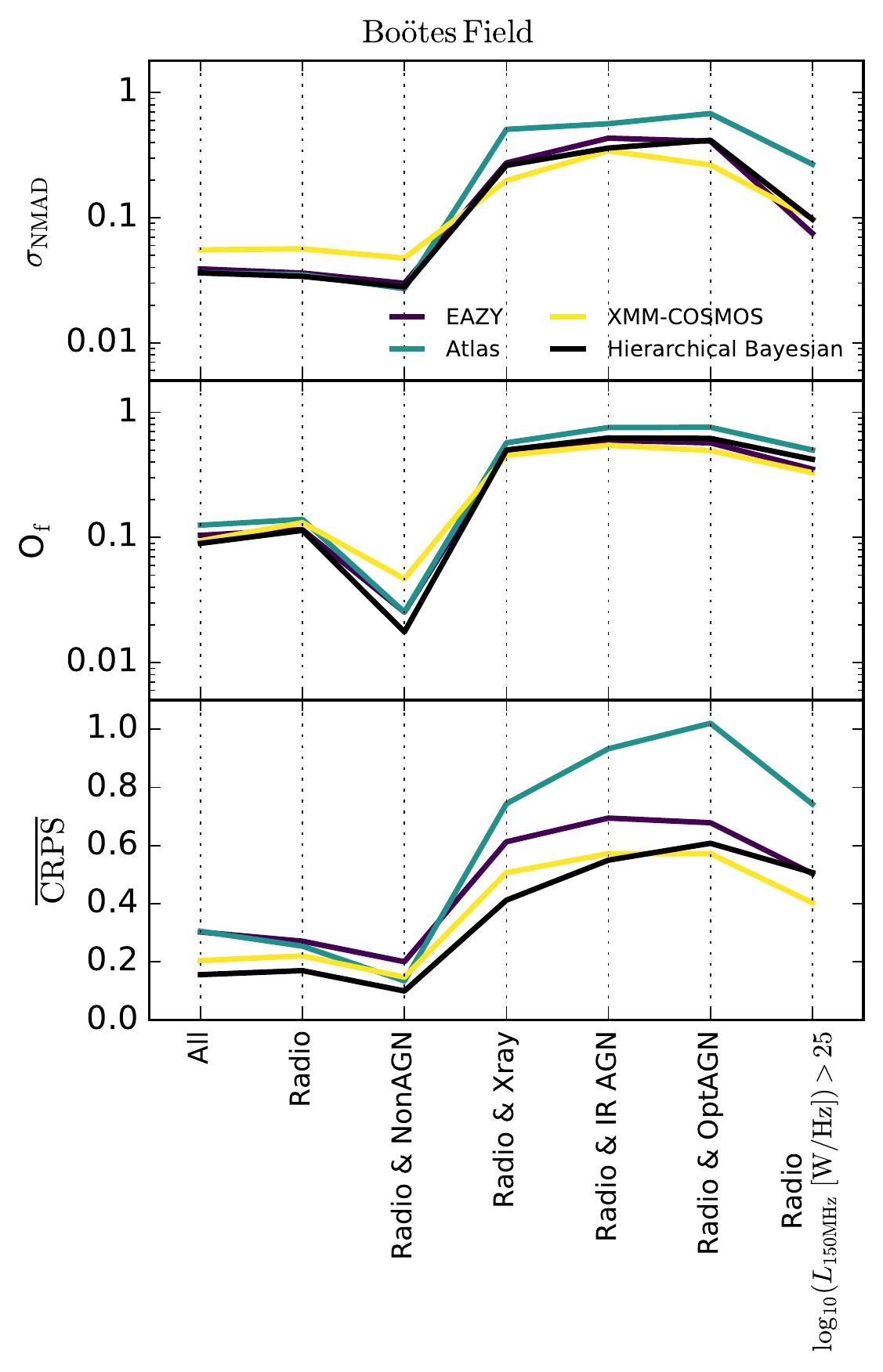}
  \caption{Visualised photometric redshift performance in three metrics ($\sigma_{\textup{NMAD}}$, $O_{\textup{f}}$, $\overline{\textup{CRPS}}$; see Table~\ref{tab:definitions}) for the different Bo\"{o}tes field radio source subsamples. }
  \label{fig:lin_stats_bootes}
\end{figure}

 \begin{figure}
 \centering
  \includegraphics[width=0.96\columnwidth]{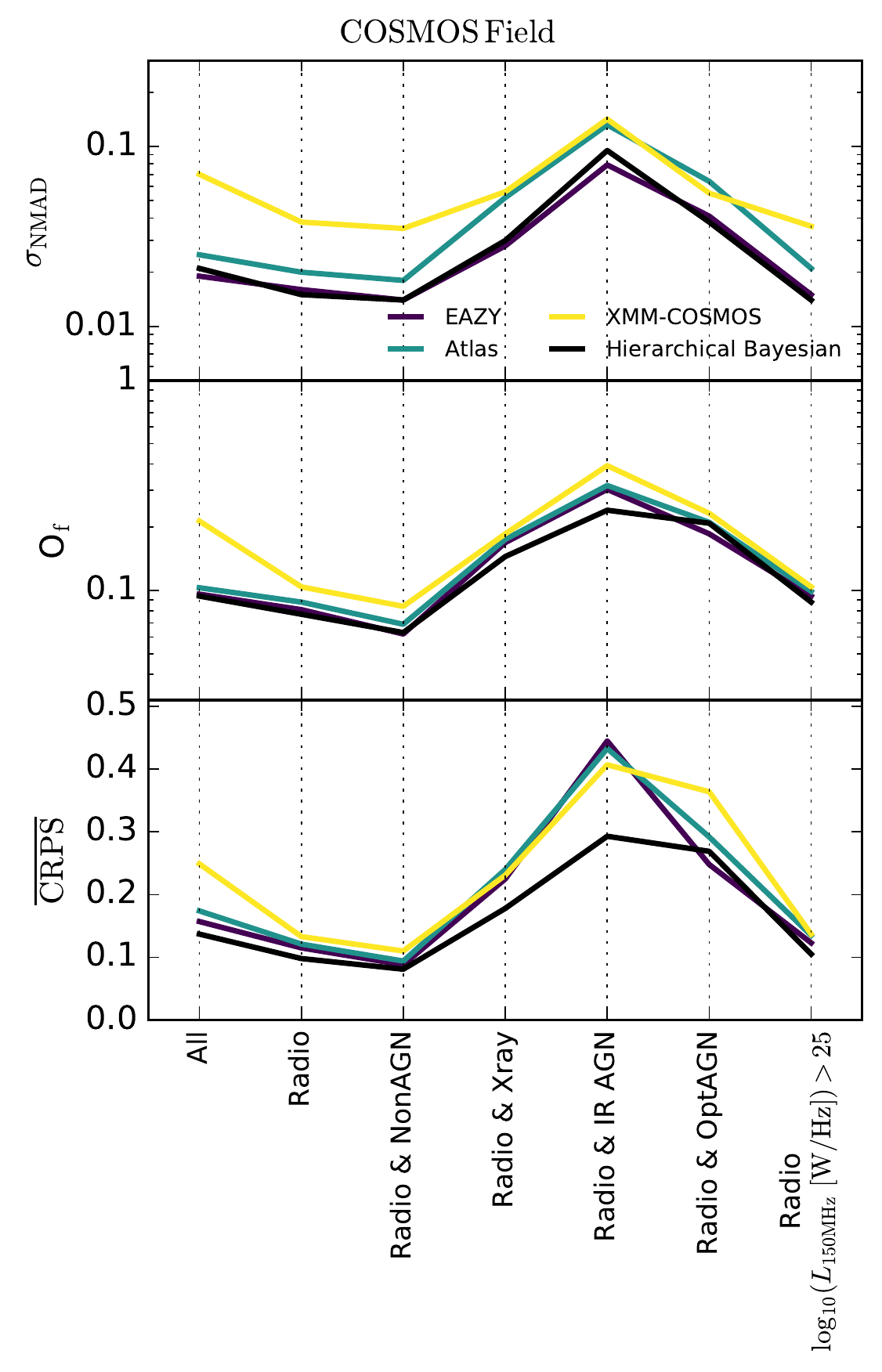}
  \caption{Visualised photometric redshift performance in three metrics ($\sigma_{\textup{NMAD}}$, $O_{\textup{f}}$, $\overline{\textup{CRPS}}$; see Table~\ref{tab:definitions}) for the different COSMOS field radio source subsamples..}
  \label{fig:lin_stats_cosmos}
\end{figure}

\begin{table}
\centering
\caption{Photometric redshift quality statistics for the derived combined consensus PDFs. The statistical metrics (see Table~\ref{tab:definitions}) are shown for the full spectroscopic sample, the radio detected sources and for various subsets of the radio population. 
}\label{tab:photzstats-hb}
\begin{tabular}{ccccccc}
Field & $\sigma_{f}$ & $\sigma_{\text{NMAD}}$ & Bias & O$_{f}$ & $\sigma_{\text{O}_{f}}$ & $\overline{\textup{CRPS}}$\\
\hline
\multicolumn{6}{c}{All Sources}\\
Bo\"{o}tes & 0.629 & 0.034 & -0.003 & 0.090 & 0.048 & 0.178 \\
COSMOS &  0.329 & 0.021 & -0.005 & 0.094 & 0.048 & 0.137 \\
\hline
\multicolumn{6}{c}{All Radio Sources}\\
Bo\"{o}tes & 0.442 & 0.034 & -0.003 & 0.111 & 0.044 & 0.174 \\
COSMOS & 0.215 & 0.015 & -0.002 & 0.077 & 0.042 & 0.098 \\

\hline
\multicolumn{6}{c}{Radio Sources - Non X-ray/IR/Opt AGN}\\
Bo\"{o}tes & 0.438 & 0.027 & -0.001 & 0.020 & 0.040 & 0.104 \\
COSMOS & 0.200 & 0.014 & -0.002 & 0.063 & 0.039 & 0.081 \\

\hline
\multicolumn{6}{c}{Radio Sources - X-ray AGN}\\
Bo\"{o}tes & 0.552 & 0.283 & -0.023 & 0.500 & 0.072 & 0.444 \\
COSMOS &  0.219 & 0.030 & 0.002 & 0.145 & 0.055 & 0.178 \\

\hline
\multicolumn{6}{c}{Radio Sources - IR AGN}\\
Bo\"{o}tes & 0.509 & 0.323 & -0.114 & 0.603 & 0.081 & 0.556\\
COSMOS & 0.389 & 0.095 & -0.001 & 0.241 & 0.071 & 0.293 \\
\hline
\multicolumn{6}{c}{Radio Sources - Opt AGN}\\
Bo\"{o}tes & 0.505 & 0.401 & -0.108 & 0.595 & 0.071 & 0.607 \\
COSMOS & 0.173 & 0.038 & -0.008 & 0.209 & 0.053 & 0.269  \\
\hline
\multicolumn{6}{c}{Radio Sources - $\log_{10}(L_{150\textup{MHz}} \textup{[W / Hz]}) > 25$} \\
Bo\"{o}tes &  0.398 & 0.122 & -0.039 & 0.412 & 0.056 & 0.525 \\
COSMOS &  0.182 & 0.014 & -0.001 & 0.088 & 0.041 & 0.105 \\

\end{tabular}
\end{table}

In Fig.~\ref{fig:lin_stats_bootes} and \ref{fig:lin_stats_cosmos} we illustrate the $\sigma_{\textup{NMAD}}$, $O_{\textup{f}}$ and $\overline{\textup{CRPS}}$  performance of the new consensus redshift estimates in each of the source population subsets (see also Table~\ref{tab:photzstats-hb}).
For the full spectroscopic samples and almost all subsets of the radio detected populations, the HB photo-z estimates either match the scatter and outlier fraction performance of the best individual template set (c.f. Table~\ref{tab:photzstats}) to within 10\% or outperform all of the estimates.
The high performance of the HB photo-z is consistent across both data sets and substantially improves upon the individual template sets in several areas.

\begin{figure}
\centering
	\includegraphics[width=0.98\columnwidth]{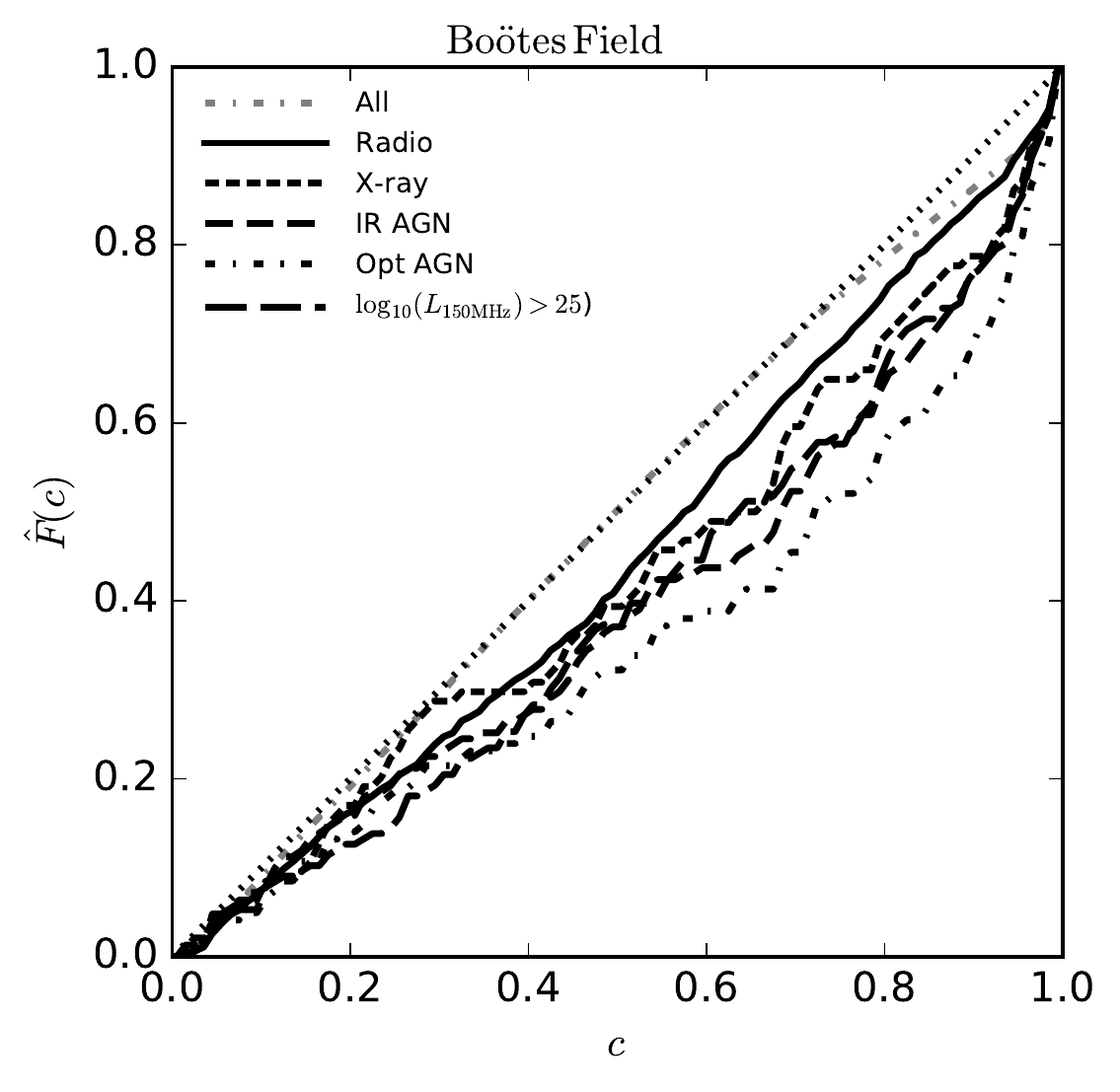}
  \caption{The $\hat{F}(c)$ plots for the consensus redshift PDF produced by the hierarchical Bayesian combination of the three different template sets. Plotted are the cumulative distributions for all LOFAR detected sources (solid line) as well the subsets of the radio sample which are X-ray detected (short dashed line), infrared AGN (medium dashed line) and high power radio sources (long dashed line).}
  \label{fig:qq_hb}
\end{figure}

\begin{figure}
\centering
	\includegraphics[width=0.98\columnwidth]{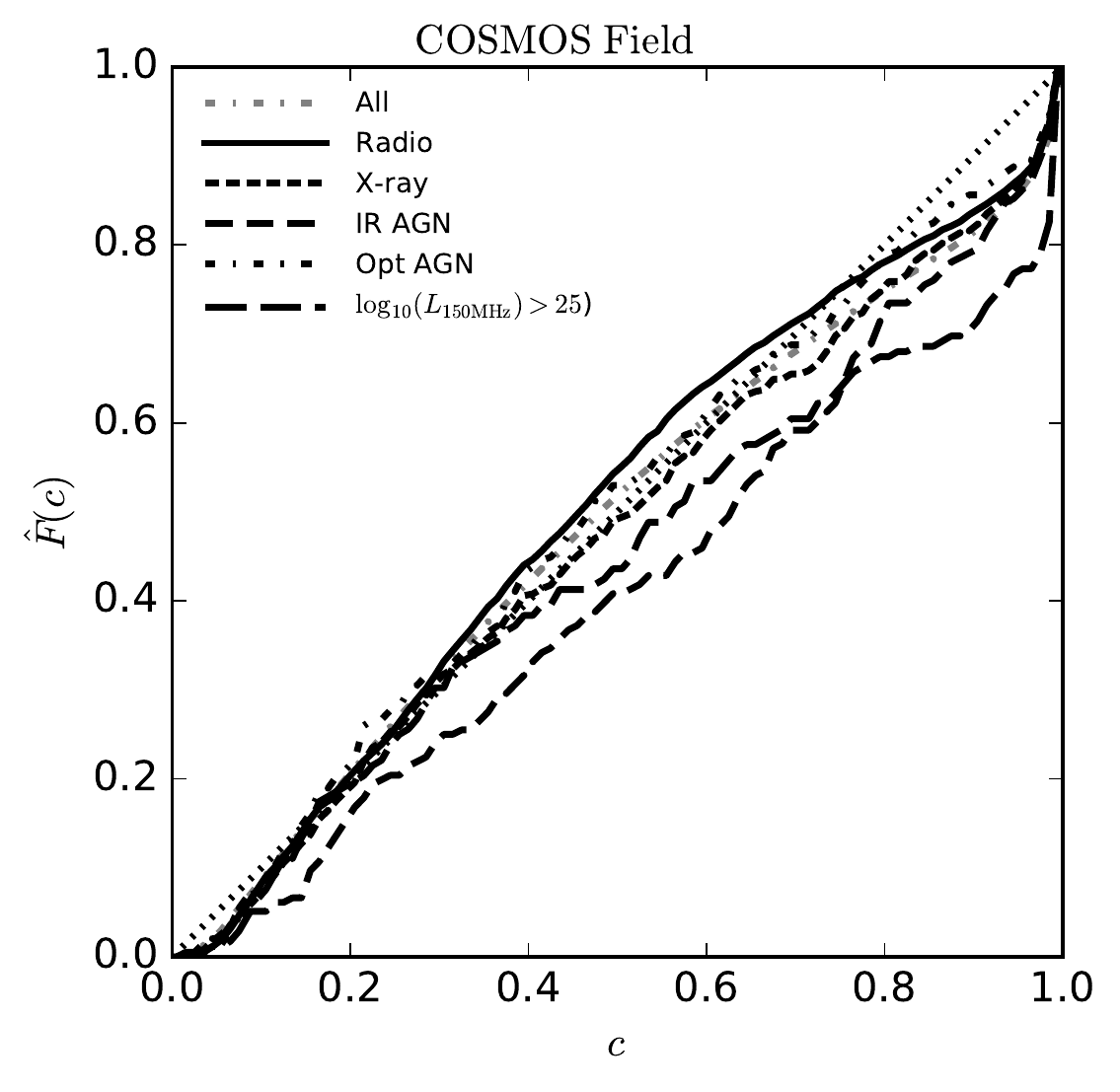}
  \caption{The $\hat{F}(c)$ plots for the consensus redshift PDF produced by the hierarchical Bayesian combination of the three different template sets. Plotted are the cumulative distributions for all spectroscopic sources (orange dash-dot line), all radio detected sources (solid line) as well the subsets of the radio sample which are X-ray detected (short dashed line), infrared AGN (medium dashed line) and high power radio sources (long dashed line).}
  \label{fig:qq_hb_cosmos}
\end{figure}

The improved performance of the consensus photo-zs also extends to the $P(z)$ distributions.
The performance of the HB redshifts in the mean continuous ranked probability score ($\overline{\textup{CRPS}}$) is significantly better than any individual redshift estimate.
In Figures~\ref{fig:qq_hb} and \ref{fig:qq_hb_cosmos} we show the $\hat{F}(c)$ distributions for the Bo\"{o}tes and COSMOS samples respectively.
For both fields, not only is the overall $P(z)$ accuracy for the full spectroscopic redshift sample improved, but the $P(z)$ accuracy for the radio detected population (and all subsets) are also improved.
Variation in $P(z)$ accuracy between the different radio subsets is significantly reduced.

\begin{figure}
\centering
  \includegraphics[width=0.99\columnwidth]{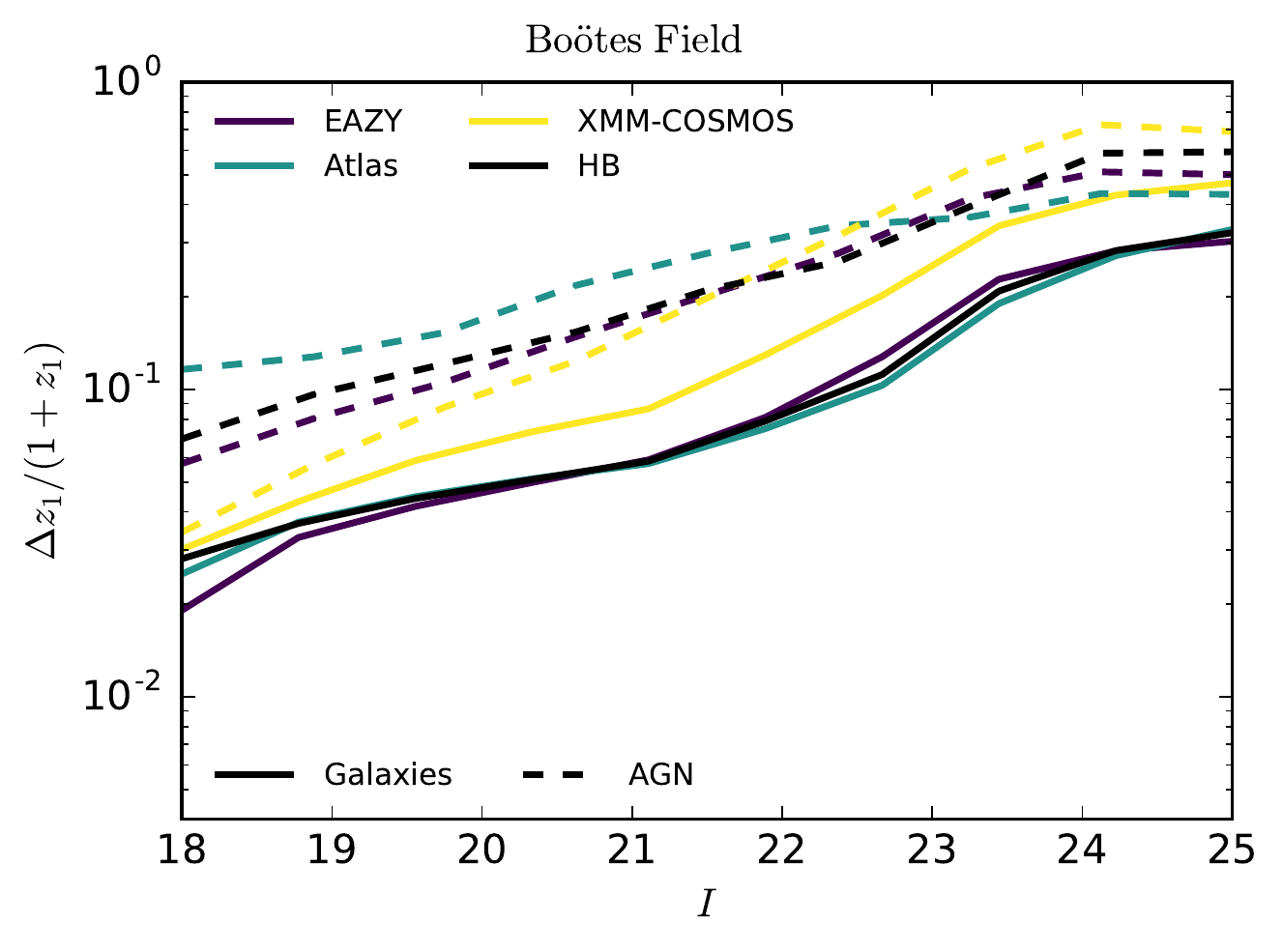}
 \includegraphics[width=0.99\columnwidth]{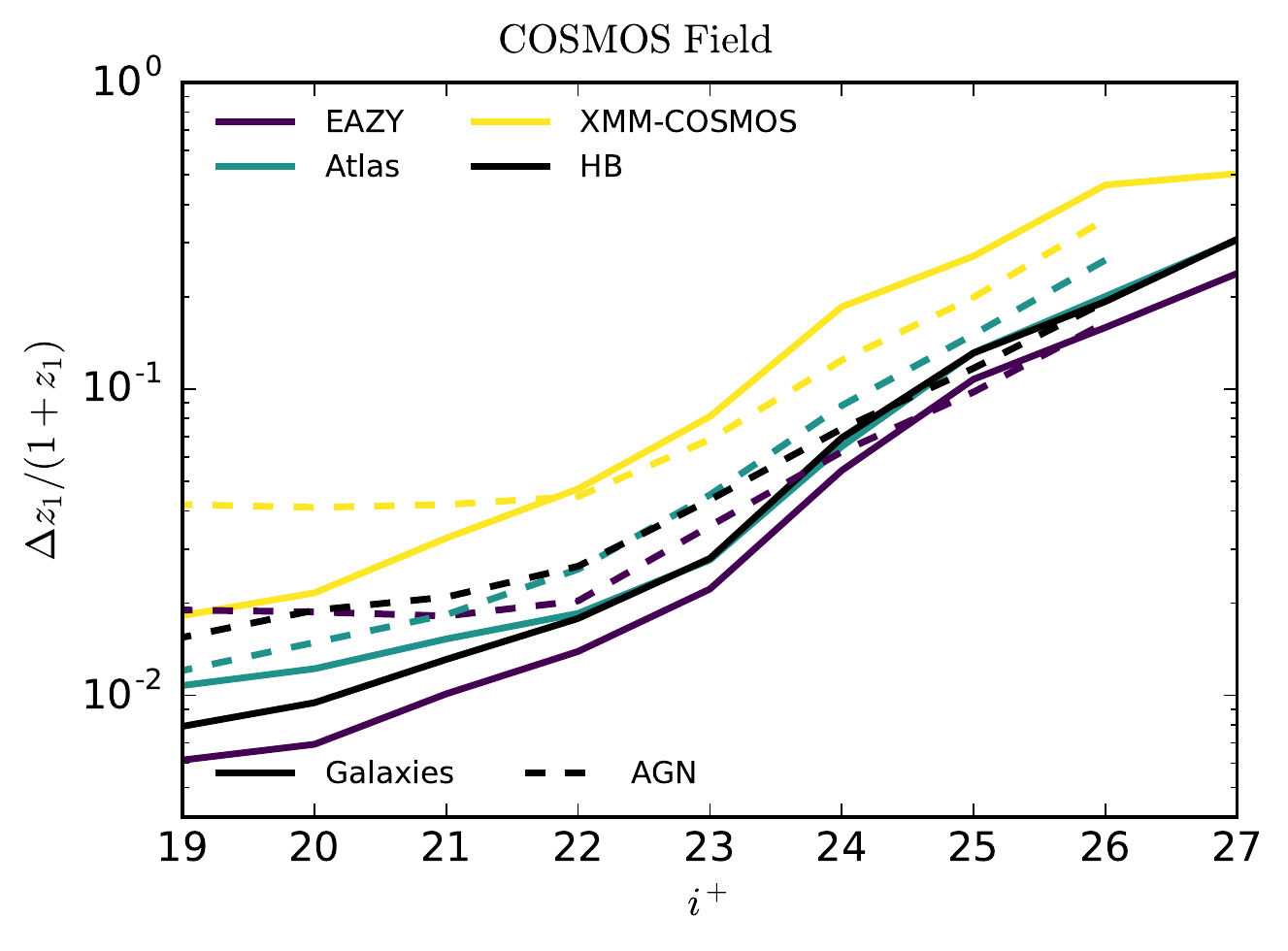}
  \caption{Median positive 80\% highest probability density confidence intervals, $\Delta_{z_{1}}$, above the primary redshift solution, $z_{1}$, as a function of $I$($i^{+}$) magnitude for both the Bo\"{o}tes (top) and COSMOS (bottom) fields. We illustrate only the upper error bounds to improve clarity by allowing a logarithmic scale. Within the primary peak, positive and negative errors are found to be very symmetrical; negative errors for each estimate follow the same magnitude trends.}
  \label{fig:sigma_err_vs_z}
\end{figure}

Finally, we note that the average uncertainty on individual source for the HB photo-z estimates remain competitive with those of the best individual template set.
In Fig.~\ref{fig:sigma_err_vs_z}, we present the median 80\% highest probability density confidence intervals, $\Delta_{z_{1}}$, around the primary redshift solution, $z_{1}$, as a function of $I$($i^{+}$) magnitude.

It is important to note here that the observed improvement in redshifts for the HB consensus photo-zs results primarily from the combining of multiple estimates and is not driven by the magnitude prior.
When folding in the magnitude priors to each individual estimate, there is only a very minor improvement in the photo-zs (namely a small reduction in outlier fraction).

\section{Discussion}\label{sec:discussion}
\subsection{Radio surveys for studying galaxy and AGN evolution}
In the preceding two sections we have presented a large amount of detailed analysis on photometric redshift estimates for two deep radio continuum surveys, including a wealth of statistics and comparisons that can be somewhat abstract.
To interpret the results presented in these sections, it is worth revisiting the questions we specifically posed in the introduction.
Firstly, how does the photometric redshift accuracy of radio sources vary as a function of redshift and radio luminosity?
And secondly, do the current methods and optimization strategies developed for `normal' galaxies or other AGN populations in optical surveys extend to radio selected galaxies?

The answer to the first question is partially provided in Section~\ref{sec:results}.
Across a wide range in radio luminosity, the measured photo-z scatter remains approximately constant, regardless of which template set is used.
In contrast, there is a much stronger evolution in the measured outlier fraction, which increases significantly between $23 < \log_{10}(L_{150\textup{MHz}}) < 27$.
At the very faintest fluxes probed by the LOFAR and VLA data used in this analysis (i.e. sub-mJy), the photo-zs for the radio source population perform very well; comparable to the overall properties of the radio undetected source population.

Based on the results of our hierarchical Bayesian combination photo-z estimates presented in Section~\ref{sec:hbmethod}, our answer to the second question posed has to be a yes.
Consensus redshifts from hierarchical Bayesian combination \emph{do} improve the redshift accuracy for these populations whilst sacrificing no accuracy for the non-radio population.
The redshifts produced perform better than any individual template set/method for both the full spectroscopic sample and the radio detected population.
The success of the ensemble redshifts is an excellent illustration that the techniques developed to provide marginal gains for the `normal' galaxy population \citep{Dahlen:2013eu,CarrascoKind:2014jg} can also provide very significant improvements for more diverse datasets.

While the consensus estimates do improve on the photo-z predictions for radio sources which also satisfy optical and IR AGN criteria (and to a lesser degree have strong X-ray flux), the overall quality of these estimates still remains very poor compared to those of the general radio source population.
\citet{Salvato:2008ef,Salvato:2011dq} illustrated that although the use AGN-dominated SEDs in the photometric redshift fitting can improve results for AGN photo-zs, additional steps are required to maximise the accuracy: namely strict magnitude priors based on optical morphology and corrections for variability in optical magnitudes between observations at different epochs.
Such steps can only be taken in a very select number of fields (in the case of variability) or in very small areas with high-resolution optical imaging (for morphology selection), making it impractical to incorporate these steps in our photo-z strategy\footnote{As optical surveys with long-term variability measurements (e.g. PanSTARRs Medium Deep Survey or the Large Synoptic Survey Telescope \citep{2012arXiv1211.0310L}) reach the depths required for deep extragalactic surveys, such corrections for variability will become significantly easier to implement and offer significant photo-z improvements for some source types.}.
However, it is also these populations have been shown to benefit from empirical (or machine learning) photo-z estimates \citep[e.g.][]{Brodwin:2006dp}.
In future, the expansion of the hierarchical Bayesian analysis to include more redshift estimates tailored for the difficult quasar populations should therefore offer further significant improvements.

Further improvements to photo-z estimates for the radio population will also be greatly aided by the forthcoming WEAVE-LOFAR spectroscopic survey \citep{Smith:2016vw}.
WEAVE-LOFAR will obtain $>10^{6}$ spectra for radio sources from the LOFAR 150 MHz survey; providing spectroscopic redshifts and source classifications for an unprecedented number of radio sources.
In particular, the deepest tier of the survey will target sources as faint as $S_{\nu, 150} \approx 100 \mu\textup{Jy~beam}^{-1}$ over several deep fields.
Such a sample will provide an extensive and unbiased training sample that can be used to improve photo-z estimates for the sub-mJy radio sources in the widest tier of the LOFAR survey and many others besides.

In addition to providing samples for photo-z evaluation/training which are more representative, significantly larger samples of spectroscopic redshifts will also offer potential improvements in the HB combination procedure outlined in this paper.
By allowing for the calibration of photo-zs and tuning of hyper-parameters in smaller subsets (e.g. split into bins in several parameters; optical magnitude/redshift/radio power etc.), further gains to the consensus redshift accuracy and precision will be gained \citep{CarrascoKind:2014jg}.

\subsection{Radio surveys for cosmology}
One of the key scientific capabilities provided by the next generation of radio interferometers like the Square Kilometre Array is as a tool for studying cosmology 
\citep{2012MNRAS.427.2079C,2014MNRAS.442.2511F, 2015aska.confE..18J}.
One such avenue for studying cosmology with radio surveys is through weak lensing (WL) experiments \citep{Brown:2015vu}.
Thanks to the different systematics offered by radio continuum observations (both in the intrinsic populations explored and the instrumentation), weak lensing studies with the SKA will be highly complementary to the increasingly powerful optical weak lensing studies planned for the next decades \citep[e.g. EUCLID;][]{2011arXiv1110.3193L}.

As with their optical counterparts, weak lensing studies with the SKA will be heavily reliant on photometric redshift estimates based on the all sky photometric data available.
While extensive effort is being invested in reaching the redshift accuracy requirements for optical weak lensing experiments \citep[e.g.]{Sanchez:2014gf,CarrascoKind:2014jg}, a key question for the SKA experiments is what effect does the radio selection have on the expected photo-z accuracies and biases?

The radio continuum depths explored in the study do not reach the faint fluxes expected for planned weak lensing studies outlined in \citet{Brown:2015vu}, so we cannot conclusively say what the full effects might be.
Nevertheless, based on the available results for the brighter radio population it is relatively clear that the prognosis for SKA WL studies remains tied to that of the comparable optical studies.

The radio source population for which our photometric redshift estimates are at least as accurate are those radio sources which are also identified as luminous X-ray sources or host dust obscured AGN.
In this regard the SKA WL experiments will be limited by the same source types as will effect the optical WL experiments.
The radio source population which are \emph{not} classified as likely AGN were found to have more accurate photometric redshift estimates than the population not detected by radio imaging.
Provided the techniques and selection criteria developed for removing the bias of AGN in optical WL experiments can be applied equally to the SKA WL samples, the radio continuum selection should not present any critical problems.
More studies will be required to test this once the SKA pathfinder and precursor surveys reach their full depths and large samples of un-biased spectroscopic redshifts are available.

\section{Summary}\label{sec:summary}
We have presented a study of template based photometric redshift accuracy for two samples of galaxies drawn from a wide area \citep[NDWFS Bo\"{o}tes;][]{Jannuzi:1999wu} and a deep \citep[COSMOS;][]{Laigle:2016ku} survey field. We calculate photometric redshifts using three different galaxy template sets from the literature.
The three template sets represent libraries which are either commonly used in photometric redshift estimates within the literature \citep{Brammer:2008gn,Salvato:2008ef} or are designed to cover the broad range of SEDs observed in local galaxies \citep{Brown:2014jd}.

Exploring the photometric redshift quality as a function of galaxy radio properties, we find:
\begin{itemize}
	\item At low-redshift ($z < 1$), radio detected galaxies typically have better photo-z scatter and outlier fractions than galaxies with comparable magnitudes, redshifts and colours but are undetected in radio. However, as redshift increases, radio-detected galaxies perform worse than their radio undetected counterparts.
	\item Within a redshift range where photo-z quality remains relatively constant, the outlier fraction of all photo-z estimates increases towards the highest radio powers (and radio flux) while scatter remains roughly the same. This trend is independent of survey field and template set.
	\item Photo-zs for radio sources not identified as AGN through X-ray, optical or IR selection criteria perform comparably to radio un-detected sources at the same redshifts.
	\item Without additional calibration, the redshift PDFs for all three template sets are overconfident; producing error estimates which are significantly underestimated.
\end{itemize}

By combining all three photo-z estimates through hierarchical Bayesian combination \citep{Dahlen:2013eu,CarrascoKind:2014jg} we are able to produce a new consensus estimate which outperforms any of the individual estimates which went into it.
The consensus redshift estimates match or better the measured scatter or outlier fraction of the best individual estimate for most radio population subsets while also providing improved predictions on the redshift uncertainties.
Nevertheless, while offering some improvement, the overall quality of photo-z estimates for radio sources which are X-ray sources or optical/IR AGN is still relatively poor; with high outlier fractions ($>20\%$) and very large scatter ($\sigma_{\textup{NMAD}}>0.2\times(1+z)$).
 Future work tailored to improving our photo-z estimates for IR/optically selected AGN will be required to achieve the some of the key science goals for deep radio continuum surveys.
In the second paper in this series we will explore the improvements offered by photometric redshift estimates from gaussian processes \citep{2016MNRAS.462..726A}, both in isolation and when combined with the template based estimates through our hierarchical bayesian combination procedure.

\section*{Acknowledgements}
The research leading to these results has received funding from the European Union Seventh Framework Programme FP7/2007-2013/ under grant agreement number 607254. This publication reflects only the author's view and the European Union is not responsible for any use that may be made of the information contained therein. KJD and HR acknowledges support from the ERC Advanced Investigator programme NewClusters 321271. PNB is grateful to STFC for support via grant ST/M001229/1. KM was supported by the Polish National Science Center (UMO-2012/07/D/ST9/02785 and UMO-2013/09/D/ST9/04030). The authors thank Mark Brodwin, Duncan Farrah, Mattia Vaccari and Lingyu Wang for valuable additional feedback and discussion. Finally, the authors thank the anonymous referee for their feedback and contributions to improving the manuscript.




\bibliographystyle{mnras}
\bibliography{library}

\bsp	
\label{lastpage}
\end{document}